\documentclass[longauth]{aa} 
\usepackage{natbib}

\usepackage{graphicx}
\usepackage{txfonts}
\usepackage{natbib}
\usepackage[colorlinks, citecolor=blue]{hyperref}

\def\kms{km\,s$^{-1}$}

\def\halpha{H${\alpha}$}

\def\hbeta{H${\beta}$}
\def\hdelta{H${\delta}$}
\def\hgamma{H${\gamma}$}
\def\l{$\lambda$}
\def\hea{\ion{He}{i}}
\def\heb{\ion{He}{ii}}

\def\snr{$S/N$}

\defcitealias{xsh1}{Paper~I}
\def\xsh{X-shooter}
\def\xshooter{X-shooter}
\def\ullyses{{\sc ULLYSES}}
\def\xshootu{XShootU}
\def\sci{{\sc science}}
\def\flat{{\sc ff}}
\def\ff{{\sc ff}}
\def\raw{{\sc raw}}
\def\flux{{\sc flux}}
\def\science{{\sc science}}

\begin{document} 
   \title{X-Shooting ULLYSES: Massive Stars at low metallicity\thanks{Based on observations collected at the European Southern Observatory under ESO program ID 106.211Z.001.}$^,$\thanks{Tables \ref{t:targets}, \ref{t:journal} and \ref{t:RV_results} are only available in electronic form at the CDS via anonymous ftp to cdsarc.u-strasbg.fr (130.79.128.5) or via \url{http://cdsweb.u-strasbg.fr/cgi-bin/qcat?J/A+A/}.}}

   \subtitle{II. DR1: Advanced optical data products for the Magellanic Clouds.}

   \author{H. Sana\inst{\ref{inst:kul}}
          \and
          F. Tramper\inst{\ref{inst:kul}}
          \and
          M. Abdul-Masih\inst{\ref{inst:eso},\ref{inst:iac},\ref{inst:ull}}
          \and
          R. Blomme\inst{\ref{inst:rob}}
          \and
          K. Dsilva\inst{\ref{inst:kul},\ref{inst:ulb}}
          \and
          G. Maravelias\inst{\ref{inst:iaasars},\ref{inst:iaforth}}
          \and
          L. Martins\inst{\ref{inst:unicid}}
          \and
          A. Mehner\inst{\ref{inst:eso}}
          \and
          A. Wofford\inst{\ref{inst:unamens}}
          \and
          G. Banyard\inst{\ref{inst:kul}}
          \and
         C.L. Barbosa\inst{\ref{inst:udfei}}
          \and
          J. Bestenlehner\inst{\ref{inst:uos}}
          \and
          C. Hawcroft\inst{\ref{inst:kul},\ref{inst:stsci}}
          \and
          D. John Hillier \inst{\ref{inst:pittpacc}}
          \and
          H. Todt\inst{\ref{inst:up}}
          \and
          C.J.K. Larkin\inst{\ref{inst:ari},\ref{inst:mpik}}
          \and
          L. Mahy\inst{\ref{inst:rob}}
          \and
          F. Najarro\inst{\ref{inst:cab}}
          \and
          V. Ramachandran\inst{\ref{inst:ari}}
           \and
          M.C. Ram\'irez-Tannus\inst{\ref{inst:mpia}}
          \and
          M.M. Rubio-Díez\inst{\ref{inst:uah}}
          \and
          A.A.C. Sander\inst{\ref{inst:ari}}
          \and
          T. Shenar\inst{\ref{inst:api}}
          \and 
          J.S. Vink\inst{\ref{inst:aop}}
          \and
F. Backs\inst{\ref{inst:api},\ref{inst:kul}}
          \and
S.~A. Brands\inst{\ref{inst:api}}
\and
P. Crowther\inst{\ref{inst:uos}}
\and
L. Decin\inst{\ref{inst:kul}}
\and
A. de Koter\inst{\ref{inst:api},\ref{inst:kul}}
\and
W.-R. Hamann\inst{\ref{inst:up}}
\and
C. Kehrig\inst{\ref{inst:gran}}
\and
R. Kuiper\inst{\ref{inst:essen}}
\and
L. Oskinova\inst{\ref{inst:up}}
\and
D. Pauli\inst{\ref{inst:up}} 
          \and
          J. Sundqvist\inst{\ref{inst:kul}}
          \and
          O. Verhamme\inst{\ref{inst:kul}}
 \and        
 the XSHOOT-U collaboration
}
   \institute{
    {Institute of Astronomy, KU Leuven, Celestijnlaan 200D, 3001 Leuven, Belgium\label{inst:kul}}
        \and
    {ESO - European Organisation for Astronomical Research in the Southern Hemisphere, Alonso de Cordova 3107, Vitacura, Santiago de Chile, Chile\label{inst:eso}}
        \and
    {Instituto de Astrofísica de Canarias, C. Vía Láctea, s/n, 38205 La Laguna, Santa Cruz de Tenerife, Spain\label{inst:iac}}
        \and
    {Universidad de La Laguna, Dpto. Astrofísica, Av. Astrofísico Francisco Sánchez, 38206 La Laguna, Santa Cruz de Tenerife, Spain\label{inst:ull}}
        \and
    {Royal Observatory of Belgium, Avenue Circulaire/Ringlaan 3, B-1180 Brussels, Belgium\label{inst:rob}}
       \and
    {Université Libre de Bruxelles, Av. Franklin Roosevelt 50, 1050 Brussels\label{inst:ulb}}
        \and  
     {IAASARS, National Observatory of Athens, 15236 Penteli, Greece\label{inst:iaasars}}
        \and 
     {Institute of Astrophysics FORTH, GR-71110, Heraklion, Greece\label{inst:iaforth}}
        \and
    {NAT - Universidade Cidade de São Paulo, Rua Galvão Bueno, 868, S\~ao Paulo, Brazil \label{inst:unicid}} 
        \and
   {Instituto de Astronom\'ia, Universidad Nacional Aut\'onoma de M\'exico, Unidad Acad\'emica en Ensenada, Km 103 Carr. Tijuana$-$Ensenada, Ensenada, B.C., C.P. 22860, M\'exico\label{inst:unamens}}
        \and
   {Centro Universitário da FEI, Dept. de Física. Av. Humberto Alencar de Castelo Branco, 3972, São Bernardo do Campo-SP, CEP 09850-901, Brazil\label{inst:udfei}}
        \and
   {Department of Physics \& Astronomy, University of Sheffield, Hicks Building, Hounsfield Road, Sheffield S3 7RH, United Kingdom\label{inst:uos}}
   \and
     {European Space Agency (ESA), ESA Office, Space Telescope Science Institute, 3700 San Martin Drive, 21218, Baltimore, MD, USA\label{inst:stsciesa}}
        \and
   {Space Telescope Science Institute, 3700 San Martin Drive, Baltimore, MD 21218, USA\label{inst:stsci}}
        \and
   {Department of Physics and Astronomy \& Pittsburgh Particle Physics, Astrophysics and Cosmology Center (PITT PACC), University of Pittsburgh, 3941 O'Hara Street, Pittsburgh, PA 15260, USA\label{inst:pittpacc}}
   \and
    {Institut f{\"u}r Physik und Astronomie, Universit{\"a}t Potsdam, Karl-Liebknecht-Str. 24/25, 14476 Potsdam, Germany\label{inst:up}}
        \and
    {Zentrum f{\"u}r Astronomie der Universit{\"a}t Heidelberg, Astronomisches Rechen-Institut, M{\"o}nchhofstr. 12-14, 69120 Heidelberg, Germany \label{inst:ari}}
       \and
    {Max-Planck-Insitut f\"ur Kernphysik, Saupfercheckweg 1, D-69117 Heidelberg, Germany \label{inst:mpik}}
       \and
   {Departamento de Astrof\'{\i}sica, Centro de Astrobiolog\'{\i}a, (CSIC-INTA), Ctra. Torrej\'on a Ajalvir, km 4,  28850 Torrej\'on de Ardoz, Madrid, Spain\label{inst:cab}}
        \and
  {Max Planck Institute for Astronomy, Königstuhl 17, 69117, Heidelberg, Germany\label{inst:mpia}}
    \and
    {Departamento de Física y Matématicas, Facultad de Ciencias, Universidad de Alcalá, Ctra. Madrid-Barcelona km 33.6, Alcalá de Henares, Madrid, Spain.\label{inst:uah}} 
   \and
    {Anton Pannekoek Institute for Astronomy, University of Amsterdam, Postbus 94249, 1090 GE Amsterdam, The Netherlands\label{inst:api}}
    \and
    {Armagh Observatory and Planetarium, College Hill, BT61 9DG Armagh, Northern Ireland \label{inst:aop}}
    \and{Instituto de Astrof\'{\i}sica de Andaluc\'{\i}a - CSIC, Glorieta de la Astronom\'{\i}a s.n., 18008 Granada, Spain\label{inst:gran}}
    \and
    {Faculty of Physics, University of Duisburg-Essen, Lotharstra{\ss}e 1, D-47057 Duisburg, Germany\label{inst:essen}}
        }
        
   
   \date{Received September 15, 1996; accepted March 16, 1997}

 
  \abstract
   {The \xshootu\ project aims to obtain ground-based optical to near-infrared spectroscopy of all targets observed by the {\it Hubble Space Telescope (HST)} under the Director's Discretionary program  \ullyses.  Using the medium resolution spectrograph \xshooter, spectra of 235 OB and Wolf-Rayet (WR) stars  in sub-solar metallicity environments have been secured. The bulk of the targets belong to the Large and Small Magellanic Clouds, with the exception of three stars in NGC~3109 and Sextans~A.
   }
   {This second paper of the series focuses on the optical  observations of Magellanic Clouds targets. It describes the uniform reduction of the UVB ($300-560$~nm) and VIS ($550-1020$~nm) \xshootu\ data  as well as the preparation of advanced data products that are suitable for homogeneous scientific analyses. }
   {The data reduction of the \raw\ data is based on  the ESO CPL \xshooter\ pipeline. We paid particular attention to the determination of the response curves. This required equal flat-fielding of the science and flux standard star data and the derivation of improved flux standard models.
   The pipeline products were then processed with our own set of routines to produce a series of advanced data products.  In particular, we  implemented slit-loss correction, absolute flux calibration, (semi-)automatic rectification to the continuum, and a correction for telluric lines. The spectra of individual epochs were further corrected for the barycentric motion, re-sampled and co-added, and the spectra from the two arms were merged into a single flux calibrated spectrum covering the entire optical range with maximum signal-to-noise ratio.}
   {We identify and describe an undocumented recurrent ghost visible on the \raw\ data. We present an improved flat-fielding strategy that limits artefacts when the \science\ and \flux\ standard stars are observed on different nights.   The improved \flux\ standard models and the new grid of anchor points allow to limit artefacts of the response curve correction on, e.g., the shape of the wings of the Balmer lines, from a couple of per cent  of the continuum level to less than 0.5\%. 
   We confirm the presence of a radial velocity shift of about 3.5~\kms\ between the UVB and the VIS arm of \xshooter\ and that there is no short term variations impacting the RV measurements. RV precision better than 1~\kms\ can be obtained on sharp telluric lines while RV precision of the order of 2 to 3~\kms\ are obtained on data with the best \snr.
   }
   { For each target observed by \xshootu, we provide three types of data products: (i) two-dimensional spectra for each UVB and VIS exposure before and after correction for the instrument response;
   (ii) one-dimensional UVB and VIS spectra as produced by the \xshooter\ pipeline  before and after response-correction, as well as after applying various processing, including  absolute flux calibration, telluric removal, normalisation and barycentric correction; and (iii) co-added flux-calibrated and rectified spectra over the full optical range, for which all available \xshootu\ exposures were combined. For the large majority of the targets, the final signal-to-noise ratio per resolution element  is above 200 in both the UVB and the VIS co-added spectra. The reduced data and advanced scientific data products are made available to the community. Together with the HST UV ULLYSES data, they should enable various science goals, from detailed stellar atmosphere and stellar wind studies, to empirical libraries for population synthesis, to study of the local nebular environment and feedback of massive stars in sub-solar metallicity environments.}

   \keywords{Atlases -- Magellanic Clouds -- Stars: early-type -- Stars: massive -- Techniques: spectroscopic }

   \maketitle
%

\begin{table*}
\centering
\caption{\xshootu\ target list, equatorial coordinates (RA, DEC), number of epochs ($N$), total exposure time ($\Sigma t_\mathrm{exp}$) and total \snr\ (per pixel  of 0.02~nm) computed in the continuum of each arm (see Sect.~\ref{ss:snr}).  The last column provides the time difference between the start of the first and last observing epochs. The full table is available at CDS.}
\label{t:targets}
\begin{tabular}{l c c c c r c c r r}
\hline
\hline      
       &    &     & \multicolumn{3}{c}{UVB arm} & \multicolumn{3}{c}{VIS arm}\\
Object & RA & Dec & $N$ & $\Sigma t_\mathrm{exp}$ & \snr & $N$ & $\Sigma t_\mathrm{exp}$ & \snr & $\Delta$MJD \\
       &(HH:AM:AS.ss) &(dDD:MM:SS.s)&      &               (s)     &      &      &               (s)   &  & (d)\hspace*{2mm}\\
\hline
2DFS-163 &  00:36:58.24 & $-$73:23:33.2 &   2 &  3310 &  208 &   2 &  3450 &  101 &    0.02\\
2DFS-999 &  00:54:32.16 & $-$72:44:35.6 &   2 &  3310 &  137 &   2 &  3450 &   93 &    0.02\\
2DFS-2266 &  01:07:14.27 & $-$72:13:47.5 &   2 &  3310 &  170 &   2 &  3450 &   99 &    0.02\\
2DFS-2553 &  01:09:21.95 & $-$73:15:41.9 &   2 &  3310 &  208 &   2 &  3450 &  105 &    0.04\\
2DFS-3689 &  01:24:31.74 & $-$73:21:49.3 &   4 &  7840 &  201 &   4 &  8120 &  105 &  214.20\\
2DFS-3694 &  01:24:34.45 & $-$73:09:08.9 &   1 &  1200 &  211 &   1 &  1270 &  106 &    0.00\\
2DFS-3780 &  01:26:35.29 & $-$73:15:16.3 &   2 &  2100 &  204 &   2 &  2240 &  131 &    0.02\\
2DFS-3947 &  01:30:37.19 & $-$73:25:14.4 &   2 &  3310 &  192 &   2 &  3450 &  118 &    0.02\\
2DFS-3954 &  01:30:43.11 & $-$73:25:04.1 &   2 &  3920 &  200 &   2 &  4060 &  114 &    0.02\\

... &... &... &... &... &... &... &... &... &... \\
\hline
\end{tabular}
\end{table*}

\section{Introduction}\label{s:intro}

Low-metallicity massive stars remain poorly understood. The lack of detailed understanding of the global, photospheric and wind properties of these massive stars, and of the physical mechanisms operating in their interiors and atmospheres directly  propagate to topics as diverse and important as the Universe's chemical enrichment in heavy elements \citep{2006ApJ...653.1145K,2014ApJ...797..123H}, the formation of high-ionization emitting gas \citep{2023Natur.617..261K}, the feedback of massive stars on the local environment \citep{2013A&A...558A.134D}, hosts galaxies and  reionization of the early Universe \citep{2018MNRAS.481.4940C}, as well as on the astrophysical interpretation of  gravitational wave detections \citep{2010ApJ...715L.138B,2020ApJ...892L...3A}. One of the reasons of the current limitations stems from the lack of a sufficiently high-quality observational database of a broad set of massive stars in various low-metallicity environments.

With typical metallicities of 0.5~$Z_\odot$ \citep{2009A&A...496..841H} and 0.2~$Z_\odot$, respectively, \citep{2003ApJ...595.1182B, 2022A&A...666A.189R}, the Large and Small Magellanic Clouds (LMC, SMC) are our closest window to the sub-solar metallicity environment. Furthermore, both galaxies are  close enough for individual massive stars to be studied in great details. In this context, the \ullyses\ Director's Discretionary program \citep{2020RNAAS...4..205R, 2023AAS...24122302R} dedicated 500 orbits of the {\it Hubble Space Telescope} ($HST$)  to obtain high-quality far-ultraviolet (FUV) spectroscopy of a representative set of OB and WR stars in the Magellanic Clouds. The initial \ullyses\ sample was further complemented by archival HST observations to reach a grand total of 130 LMC and 124 SMC targets.

\begin{table*}
\centering
\caption{Journal of the observations. The first column gives the target name. Columns 2 to 4 (resp.~5 to 7) give  the modified Julian date (MJD) at the start of the exposure, the exposure time and the (per pixel) \snr\ computed in the continuum (see Sect.~\ref{ss:snr}) for the UVB (resp.~VIS) arm of the rebin individual spectra ($\Delta \lambda=0.02$~nm; see Sect.~\ref{ss:coadd}). Columns 8 and 9 provide the name of the \flux\ standard star used to calibrate the response curve and the time difference ($\Delta t$) between the start of the \science\ and \flux\ observations. The last column contains a flag for additional comments provided in the online version of the table. The full table is available at CDS.}
\label{t:journal}
\begin{tabular}{l c c r c c r l c l}
\hline
\hline
       & \multicolumn{3}{c}{UVB arm}& \multicolumn{3}{c}{VIS arm}\\
Object & MJD & $t_\mathrm{exp}$ & \snr &  MJD & $t_\mathrm{exp}$ & \snr & Flux STD & $\Delta t$ & Comment \\
       & & (s) & & & (s) & & & (d)\\
\hline
2DFS-163 & 59150.059 &  1655 &  139 & 59150.059 &  1725 &   78 &  LTT3218   &  $-$0.3\\
          & 59150.078 &  1655 &  166 & 59150.078 &  1725 &   70 &  LTT3218   &  $-$0.3\\
2DFS-999 & 59157.207 &  1655 &   112 & 59157.207 &  1725 &   71 &  LTT7987   &   $+$0.2\\
         & 59157.227 &  1655 &   106 & 59157.227 &  1725 &   64 &  LTT7987   &   $+$0.2\\
2DFS-2266 & 59179.160 &  1655 &  125 & 59179.160 &  1725 &   72 &  GD 71  & $-$0.1\\
           & 59179.180 &  1655 &  129 & 59179.180 &  1725 &   73 &  GD 71  & $-$0.1\\
2DFS-2553 & 59161.117 &  1655 &  141 & 59161.117 &  1725 &   70 &  LTT3218   & $-$0.2\\
          & 59161.160 &  1655 &  145 & 59161.160 &  1725 &   84 &  LTT3218   & $-$0.2\\
2DFS-3689 & 59186.145 &  1960 &  145 & 59186.145 &  2030 &   63 &  FEIGE110   &  $+$0.1\\
          & 59186.168 &  1960 &  115 & 59186.168 &  2030 &   60 &  FEIGE110   &  $+$0.1\\
          & 59400.316 &  1960 &   78 & 59400.316 &  2030 &   45 &  EG274   &   $+$0.2\\
           & 59400.340 &  1960 &   72 & 59400.340 &  2030 &   49 &  EG274   &  $+$0.3\\
2DFS-3694 & 59147.297 &  1200 &  211 & 59147.297 &  1270 &  106 &  LTT3218   & $-$1.1\\
2DFS-3780 & 59195.172 &  1050 &  165 & 59195.172 &  1120 &  103 &  FEIGE110   &  $+$0.2\\
          & 59195.188 &  1050 &  154 & 59195.188 &  1120 &   87 &  FEIGE110   &  $+$0.2\\
2DFS-3947 & 59175.199 &  1655 &  147 & 59175.199 &  1725 &   83 &  EG21  &   $+$0.1\\
          & 59175.219 &  1655 &  154 & 59175.219 &  1725 &   82 &  EG21  &   $+$0.1\\
2DFS-3954 & 59174.191 &  1960 &  153 & 59174.191 &  2030 &   70 &  FEIGE110   &   $+$0.2\\
          & 59174.215 &  1960 &  162 & 59174.215 &  2030 &   90 &  FEIGE110   &   $+$0.2\\
... &... &... &... &... &... &... &... &... \\
\hline
\end{tabular}
\end{table*}

The {\it X-Shooting ULLYSES project} (\xshootu) is an accompanying ESO Large Program that aims to provide ground-based spectroscopy of all ULLYSES targets \citep[][hereafter \citetalias{xsh1}]{xsh1}. Using the ESO \xshooter\ spectrograph \citep{2011A&A...536A.105V}, \xshootu\ complements the FUV {\it HST\/} observations with homogeneous intermediate-spectral resolution spectroscopy, covering the  optical and near-infrared (NIR) domains. 
While a subset of the ULLYSES targets had existing optical spectroscopy prior to \xshootu, the spectral coverage and data quality was far from homogeneous. 

Optical spectra are indeed needed to enable the full potential of the {\ullyses} campaign.  Indeed, stellar parameters such as surface gravity ($\log g$) or projected rotation rates ($\varv \sin i$) are difficult to reliably constrain from FUV spectra alone, which translates into significant uncertainties on, e.g., stellar wind parameters and surface abundances. These, however, are important quantities to constrain theories of stellar structure,  evolution and outflows which constitute some of the most fundamental  building blocks of our current understanding of the multi-messenger Universe.

Similarly, high-quality libraries of massive star  spectra and the  physical parameters of the associated stars, are crucial ingredients for population synthesis tools that are used to  analyse distant populations of massive stars observed in integrated light. Unfortunately, empirical high-quality sub-solar metallicity libraries have been  missing. These are, however,  needed in the context of a new generation of telescopes that will probe the distant, low-metallicity universe, such as  the {\it James-Webb Space Telescope (JWST)} and upcoming 30m-class telescopes.

The goals of the \xshootu\ project have been described in \citetalias{xsh1}, including a top-level description of the target sample and the observing campaign. In this paper, we present the uniform reduction and calibration of the optical X-shooter data of the Magellanic Clouds stars. We provide advanced data products that can be used for numerous science applications, from detailed atmospheric analyses to the creation of stellar template libraries. The NIR data reduction requires specific attention and will be presented in a subsequent paper in the \xshootu\ series.

A note on the naming convention used in this paper; while Table 1 of \citetalias{xsh1} provides all target names, these are often not suitable for electronic filenames (e.g. they  contain spaces or special characters). For this reason and for consistency with the FUV {\it ULLYSES} dataset, we opted to use the MAST target names for all fits files in the \xshootu\ data release.  An overview of \citetalias{xsh1} names and corresponding MAST target names is given in Table~\ref{t:names}.

This paper is organised as follows. Section~\ref{s:obs} provides a detailed description of the observing campaign  and of the collected \raw\ data. Section~\ref{s:DR} describes the data reduction process, including non-standard procedures for the flat-fielding and instrument response correction. Slit-loss correction, absolute flux calibration, normalisation, telluric correction, resampling and co-addition of the final science products are described in Sect.~\ref{s:ADP}. Section~\ref{ss:RV} provides the radial velocities, while Sect.~\ref{s:datarelease} describes the advanced data products that are made available to the community. 

\section{Observation campaign}\label{s:obs}

\subsection{Instrumental setup} \label{ss:OBSoverview} %
The ESO \xshooter\ spectrograph uses two dichroics located after the focal plane. These dichroics split the light into three arms (UVB, VIS, and NIR), with a separation between the arms at $\lambda \approx 560$~nm and $1024$~nm \citep{2011A&A...536A.105V}. In this paper, we focus exclusively on the spectra obtained with the UVB and VIS arms. Both arms are equipped of 4k $\times$ 2k detectors, covering 12 and 15 spectral orders, respectively.

The \xshooter\ spectral resolving power ($R=\lambda / \Delta \lambda$) is set by the width of the entrance slits. The \xshootu\ campaign used slits of 0\farcs8 for the UVB arm and 0\farcs7 for the VIS arm, corresponding to $R\approx  6700$ and 11\,400.  In this setup, the point spread function (PSF) along the dispersion direction is well sampled, with 5 and 4.4 \raw\ pixels across the full-width at half-maximum of the UVB and VIS arms PSFs,  respectively. Similarly,  the UVB and VIS detectors provide a \raw\ pixel scale of $\approx$1\farcs6~pix$^{-1}$ in the spatial direction. The slow, high-gain readout mode was chosen for both detectors with no binning (readout mode: {\tt 100,1x1,hg}; gain $\approx0.63$~e$^{-}$~ADU$^{-1}$), delivering a low readout noise of 2.4~e$^{-}$. The  dark current level is extremely low ($<2$~e$^{-}$~pix$^{-1}$~h$^{-1}$) and will be neglected. Both UVB and VIS detectors have excellent cosmetics.

Each \xshootu\ target was observed in {\sc stare} mode using sequences of observation (OBs) composed of one exposure or, for the fainter objects, two back-to-back exposures with identical exposure times. The individual exposure times vary between 100~s and 2955~s according to the brightness of the stars, with the aim to obtain a signal-to-noise ratio per resolution element (\snr) of at least 100 at 490~nm and 600~nm for all LMC and SMC targets. 
No dedicated night-time calibrations were requested by the \xshootu\ program, such that we rely on the standard \xshooter\ calibration plan. The latter typically offers one flux standard star and one telluric standard star observed during the same night as the scientific observations.

An overview of the observing campaign is provided in Table~\ref{t:targets} while the full journal of observations is in Table~\ref{t:journal}. 
The observations were spread on 102 different nights from October~2020 to January~2022 and include 114 LMC and 112 SMC stars. In total, 358 UVB and 357 VIS individual exposures were obtained. 73 objects have multi-epoch observations (i.e., two or more exposures), of which 34 have these observations spread over different nights. The main reasons for these repeated observations over two nights or more, are due to operators declaring that the weather conditions were below requirements or because part of the spectrum was saturated. These repeated observations provide a limited insight into the variability of these objects.

\begin{figure*}
    \centering
    \includegraphics[width=17cm]{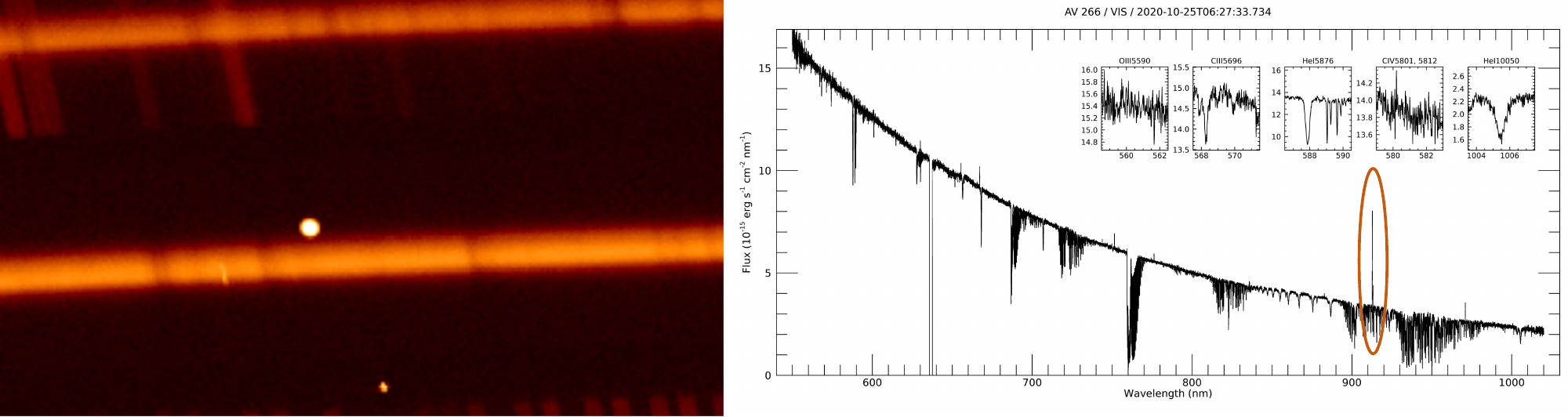}
    \caption{Example of the ghost artefact identified in various 2D \raw\ images (left panel). Whenever the ghost position overlaps with the extraction region of the science target, it produces an artefact in the extracted 1D spectrum (right panel, at $\lambda \approx910$~nm in this example). Inserts have the same units as the main panel.}
    \label{f:ghost}
\end{figure*}

\subsection{Quality control of the \raw\ data} \label{s:QC} 
We inspected each \raw\ data frame and checked for (near) saturation ($>$60\,000~ADU). We identified  one target with very strong saturation (VFTS~482) and 21 other targets where the \hbeta\ emission line or nebular lines were saturated. For those targets, observations were repeated with reduced exposure times to record the full spectrum without saturation. All data where however reduced in the same way and saturated parts of the spectra were masked in the co-addition (see Sect.~\ref{ss:coadd}).

\begin{figure}
    \centering
    \includegraphics[width=\columnwidth]{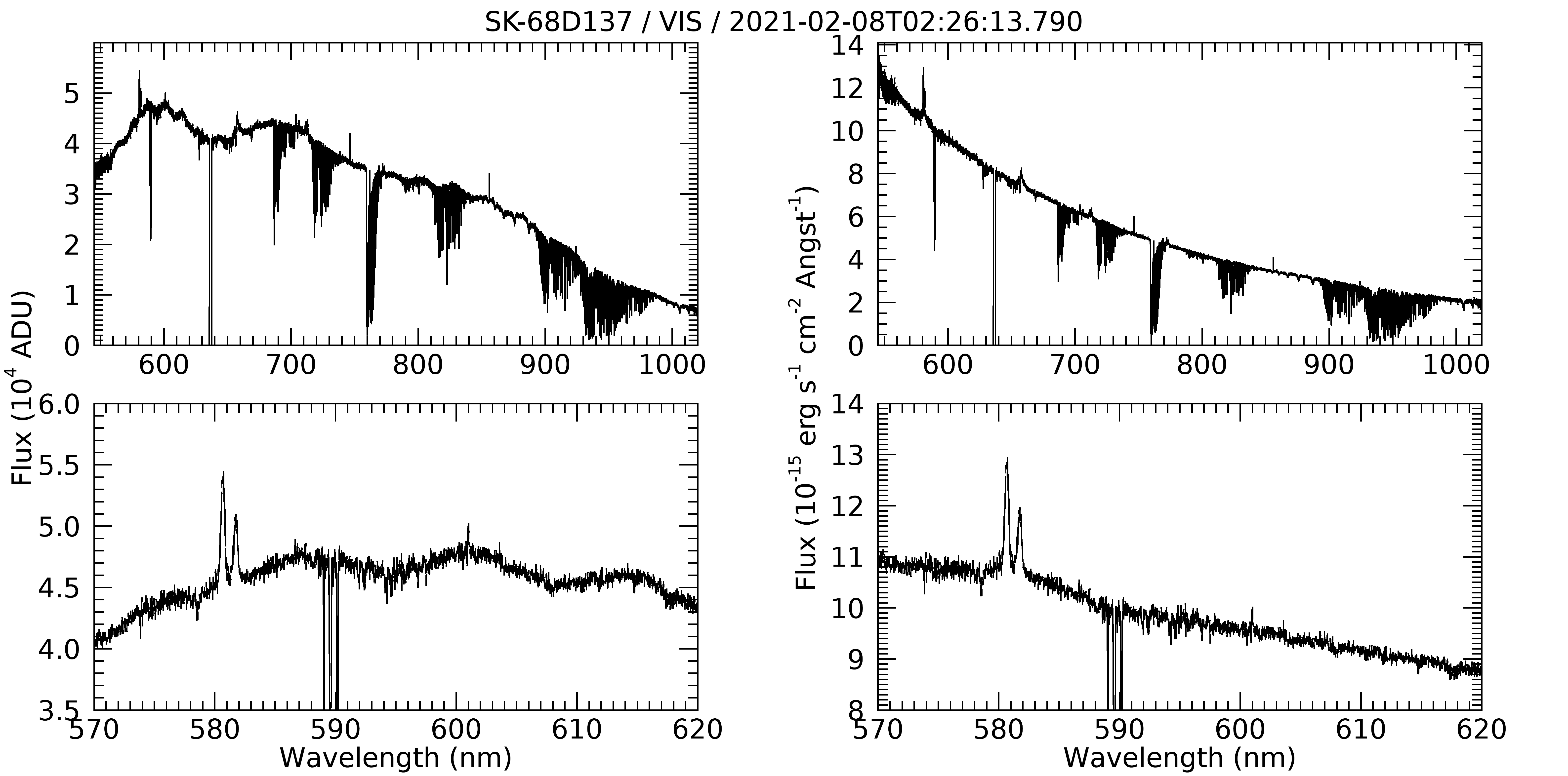}
    \caption{Comparison of a reduced spectrum of SK-68D137 without (left panels) and with (right panels) response-curve correction.}
    \label{f:RespCurve}
\end{figure}
During this quality control process, we identified the presence of an undocumented ghost that affects about 15\%\ of our observations. The ghost appears as a bright, circular spot of typically 7~pix in diameter (Fig.~\ref{f:ghost}, left panel). In most cases, the spot appears on an unused part of the detector (outer part, inter-order regions). However, in 3\%\ and 1\%\ of the frames, it landed, respectively, on a region used for the sky or overlapped with the science spectrum. In the latter case, the spot produces a clear artefact in the reduced 1D spectrum  (Fig.~\ref{f:ghost}, right panel).  In most cases, the ghost is saturated, so that it can easily be identified in the cumulative distribution of counts in each pixel. However, we have also visually identified frames where the ghost was not saturated, making it more difficult to automatically flag. 
The origin of the ghost has not been identified. 
Some back-to-back exposures separated by  10 to 15~min reveal that the ghost moves on short timescale or even disappears. In a handful of  cases, multiple ghosts could be identified in the \raw\ image.

\section{Data reduction}\label{s:DR}

\subsection{General overview } \label{ss:DRoverview}

The data reduction was performed using the ESO X-shooter pipeline v3.5.0 \citep{2011AN....332..227G}. 
The following steps were applied using dedicated routines of the pipeline: bias subtraction, order guess position estimate, order tracing, flat-fielding, wavelength calibration, spectral extraction, sky subtraction and response-curve correction. No dark correction was applied given the very low dark current of the UVB and VIS detectors.  We decided early on to work with response-curve corrected spectra as these provide better correction for the blaze functions and an overall smoother spectrum, which eases the rectification to the continuum (Fig.~\ref{f:RespCurve}). For the vast majority of the observations, \raw\ calibration frames obtained in the mornings after each observation night were adopted. When these were not available, suitable  calibration frames nearest in time were used.  

While the general data reduction strategy follows closely the standard recommendation of ESO, the following sections detail issues that were identified and mitigation strategies implemented. This section focuses on the data reduction with the ESO CPL pipeline recipes. Additional calibration steps, such as slit-loss correction, absolute flux calibration, telluric correction, and normalisation, will be addressed in the next section. 

\begin{figure}
    \centering
    \includegraphics[width=\columnwidth]{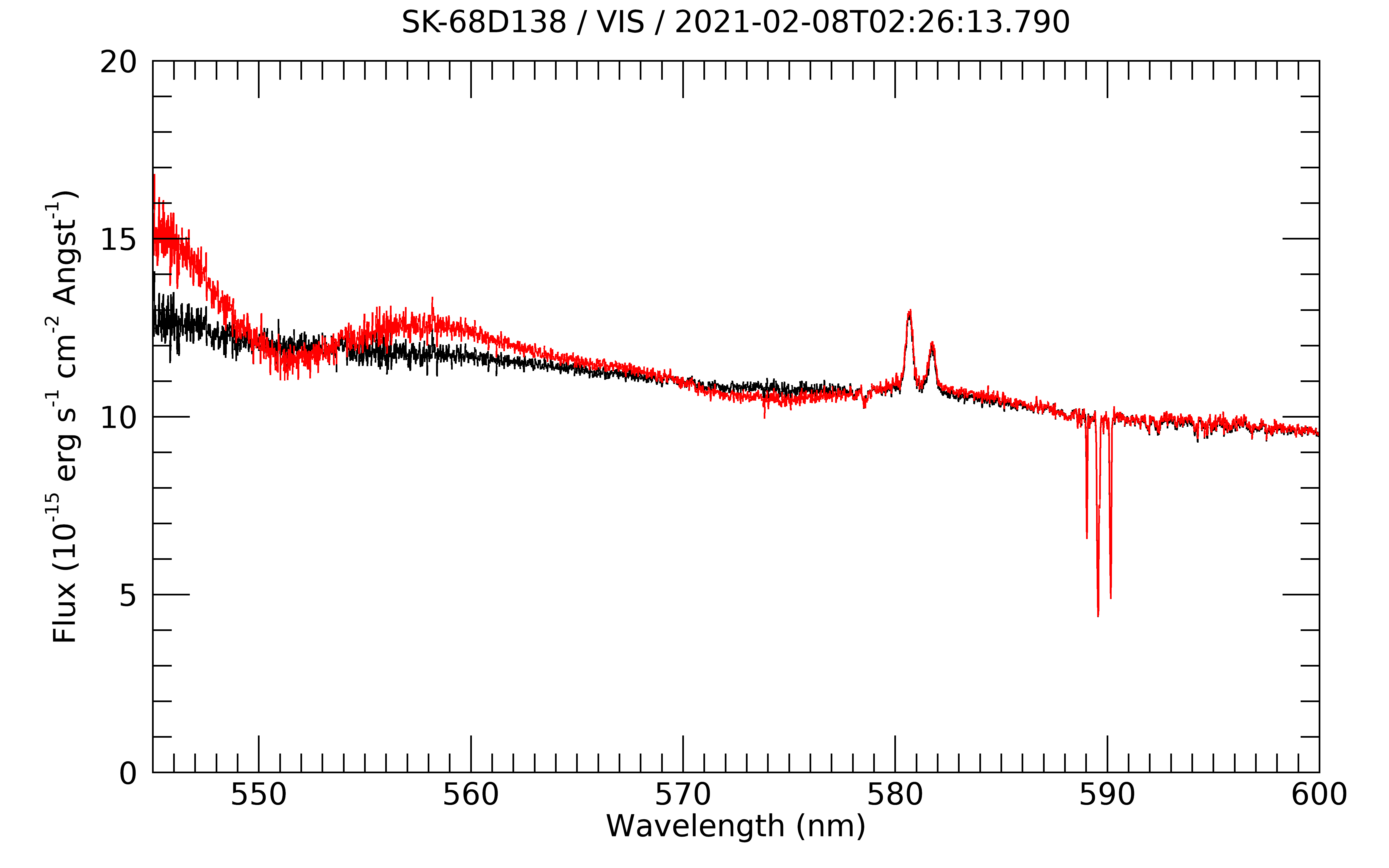}
    \caption{Comparison of reduced spectra of SK-68D137 obtained with the different flat-fielding (\flat) strategies in case the \sci\ target and the \flux\ standard star have been observed on different nights. Red: Different \flat\ frames were used to reduce the observations of the \sci\ and  \flux\ stars. Black: The same \flat\ frames were used.}
    \label{f:FF}
\end{figure}

\begin{figure}
    \centering
    \includegraphics[viewport=28 28 538 538,width=\columnwidth]{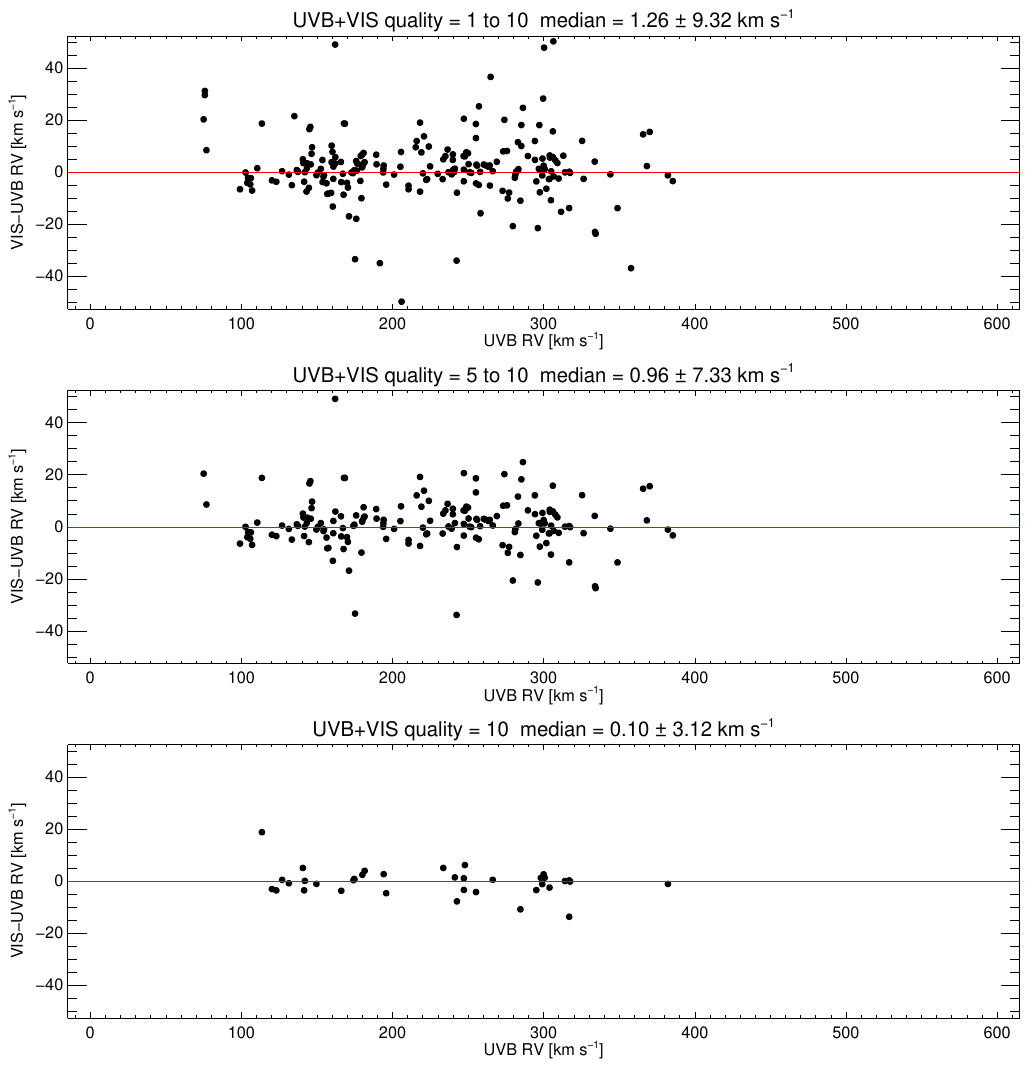}
    \caption{Radial velocity (RV) difference measured between the UVB and VIS arms after correction for known wavelength calibration offsets (Sect.~\ref{ss:wlc}). The panels show the values for different data quality (increasing from top to bottom). 
  The UVB+VIS quality of a star is given by the sum of all points listed in Table~\ref{table RV quality}.
     The median and 68.3\%\ quantile are listed in the top of each plot. 
    }
    \label{f:RV}
\end{figure}

\subsection{Flat fielding}\label{ss:flats} 

Flatfield frames (\flat)  are used in the \xsh\ calibration chain to correct for the pixel-to-pixel response variation and to determine the blaze functions of the  \'echelle orders. Under the standard \xshooter\ calibration plan, the validity of these \flat\ calibration frames is of three days and they are thus not taken daily. The default reduction procedure further uses \flat-frames with matching slit widths to reduce the \sci\ and \flux\ observations (i.e.,  0.8\arcsec/0.7\arcsec\ for the UVB/VIS \sci\ frames in our case, and  5\arcsec\ for \flux\ frames).

While this is a standard approach, the net result is that different \flat-frames are used to reduce the  \flux\ and the \sci\ observations, so that any difference in the pixel-to-pixel or blaze-correction between the \flux\ and the \sci\ data reduction leaves a residual imprint in the flux-calibrated spectrum. On average, the effect is negligible but might become noticeable for some orders close to the edge of the detectors, especially when the \ff-frames used to reduce the \flux\ and \sci\ observations are taken on different days (Fig.~\ref{f:FF}).

To investigate the most suitable flat-fielding strategy, we compared different approaches. In addition to the standard approach described above, we used the same broad-slit \ff\ frames to reduce both the \flux\ and the \sci\ frames. We also used the narrow-slit \ff\ frames corresponding to the \sci\ observations to reduce the \flux\ data. The size of the slit has a negligible effect on the outcome of the procedure. However, using the same \ff\ frames to reduce the \flux\ and \sci\ observations improves the determination of the blaze function (Fig.~\ref{f:FF}). Our final approach used the broad-slit 5\arcsec\ \flat\ frames obtained for the \flux\ observations to reduce both the \flux\ and the \sci\ frames. From our tests, using \ff-corrections obtained with different slit widths have no significant impact as long as the the \ff-slitwidths is at least as large as the smallest one used for either \flux\ or \sci. This is logical as the pixel-to-pixel sensitivity correction is performed at the pixel level on the 2D images. However, forcing the \flux\ and \sci\ frames to be corrected by the same \flat-calibration 
 guarantees that the imprints of the \flat\ correction residuals of the \flux\ and \sci\ frames cancels  out when correcting for the response curve of the instrument (see Sect.~\ref{ss:absfluxcal}).

\begin{figure*}
    \centering
    \includegraphics[width=17cm]{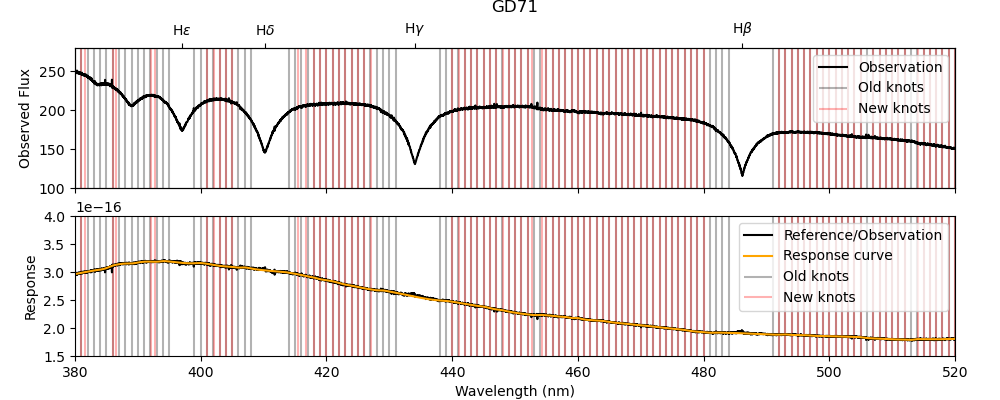}

    \caption{ Comparison of knot points between the original knots (plotted in grey) and the new knots (plotted in red) used for interpolation in determining the response curve for GD~71 in the UVB arm. The top sub-panel shows the observed spectra as a function of wavelength for one observation, while the bottom sub-panel shows the reference model divided by the observation in black and the resulting response curve after interpolation in orange.}
    \label{f:FluxModel1}
\end{figure*}

\begin{figure}
    \centering
    \includegraphics[width=\columnwidth]{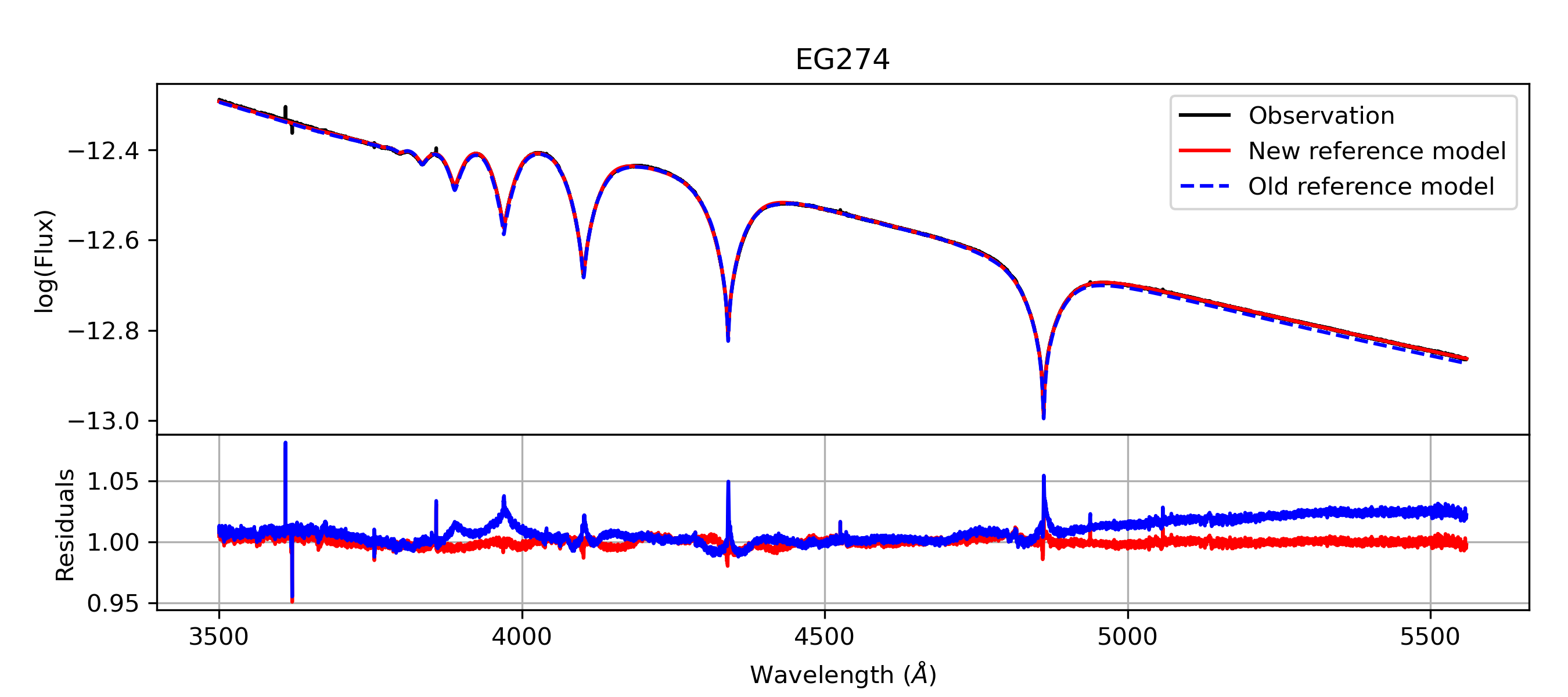}
    \includegraphics[width=\columnwidth]{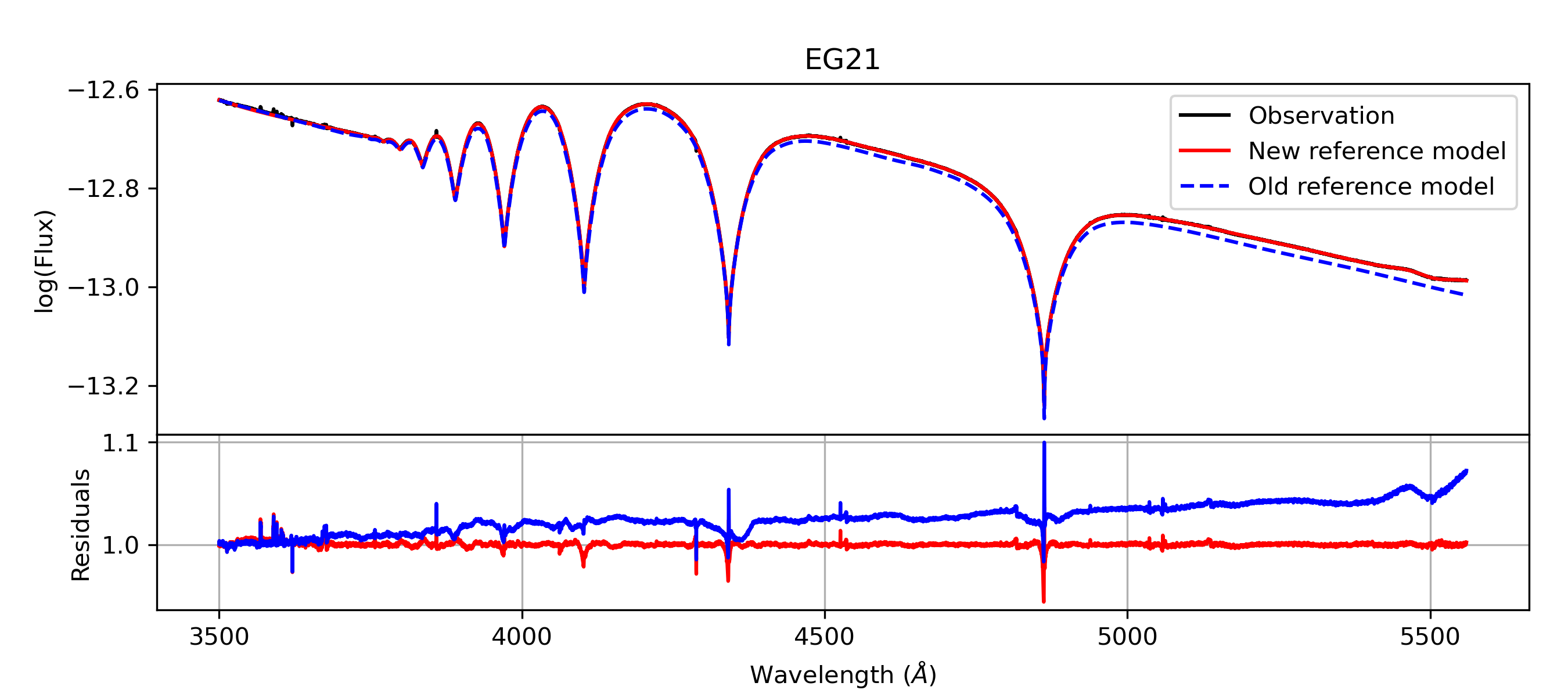}

    \caption{Comparison of new and old reference models for  the EG~274 and EG~21 \flux\ standard stars used by the \xshootu\ program.  The top sub-panel of each star show the observed co-added spectrum (in black) and the new and old reference models (in red and blue, respectively).  The bottom sub-panels show the residuals between the observation and each of the reference models using the same color scheme.}
    \label{f:FluxModel}
\end{figure}

\begin{figure}
    \centering
    \includegraphics[width=\columnwidth]{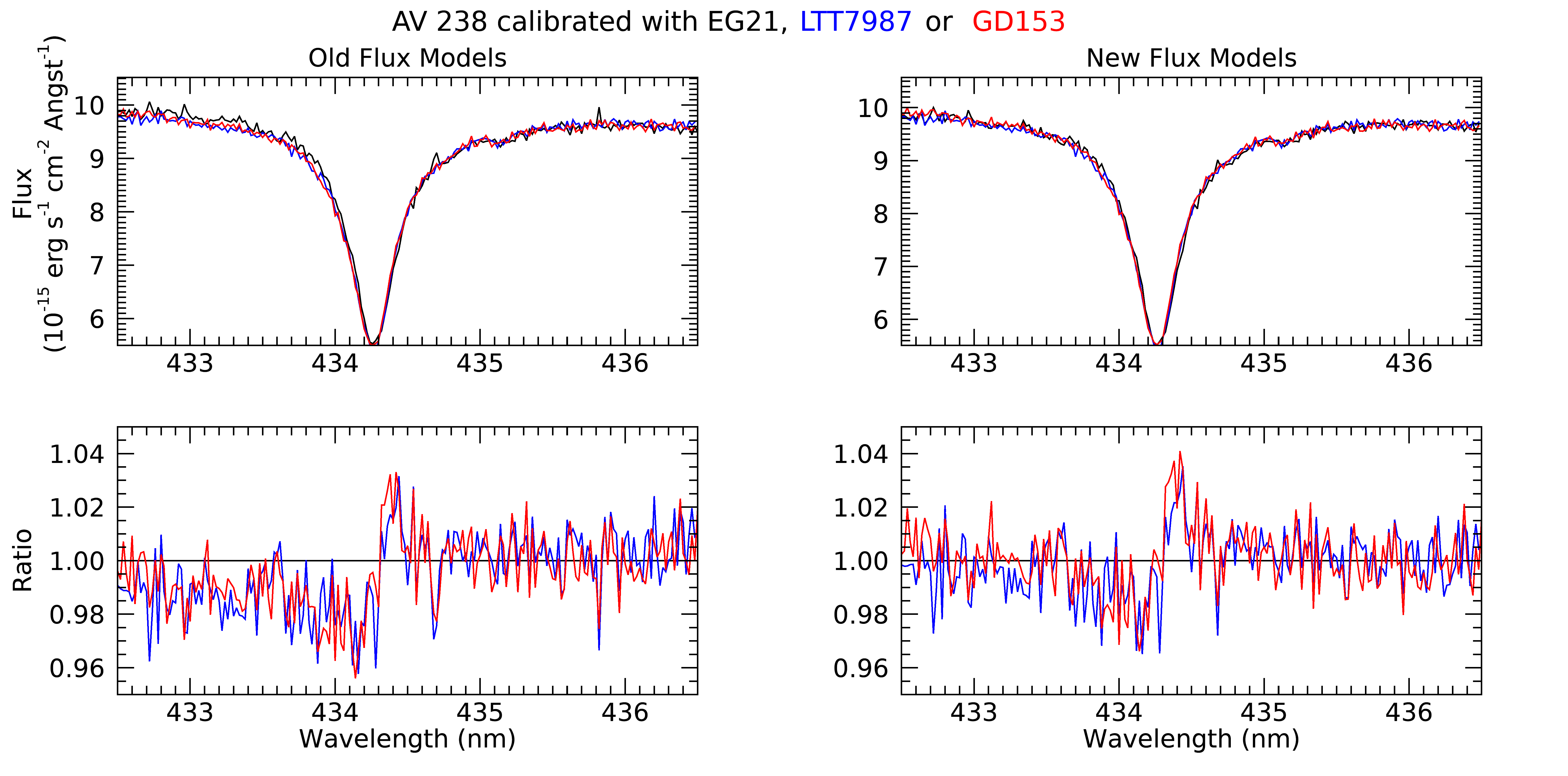}
    \includegraphics[width=\columnwidth]{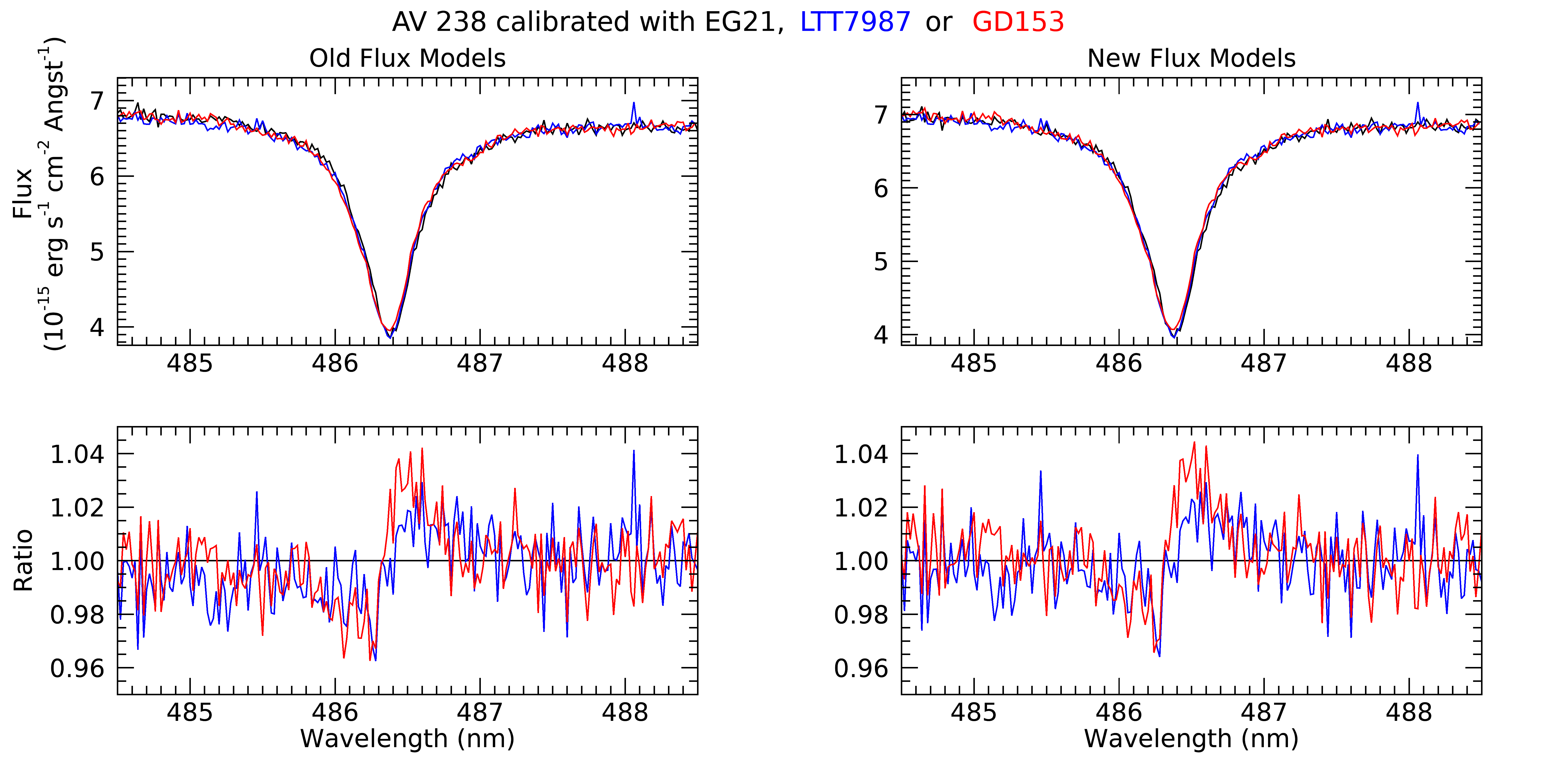}
    \caption{Comparison of the  \hgamma\ (top) and \hbeta\ (bottom) line profile of the target AV~238 for three different observations (Dec~2020, May 2021, and Jun 2021) and calibrated with three different \flux\ standard stars black: EG~21, blue: LTT~9789, red: GD~153). The residuals plots show the ratio between data calibrated with LTT~7987 (blue) and GD~153 (red) and those calibrated with EG~21. The 1 and 2\%\ inconsistencies in the blue wing of \hbeta\ and \hgamma\, respectively, have disappeared when using the new flux models. The remaining differences located in the core of the lines and is are likely of  astrophysical origin (e.g., line profile or RV variations, or nebular correction residuals).}
    \label{f:FluxModel_hbeta}
\end{figure}

\subsection{Wavelength calibration - RV shifts between arms} \label{ss:wlc}

The wavelength calibration performed by the standard ESO pipeline uses a physical model \citep{2010SPIE.7737E..1GM}. The transformation from pixel to wavelength space is optimized through the analysis of multi(9)-pinhole mask ThAr (UVB, VIS) lamp frames, instead of ThAr observations using the science slits.
The predicted positions of the lines obtained through the multi(9)-pinhole mask are fitted using a two-dimensional Gaussian to recover the actual positions on the frame.  The resulting wavelength calibration is slightly offset in wavelength compared to observations through the science slits as different optical components are in the light path. These offsets can be recovered using quality control information of ThAr slit observations, present in the ESO QC1 database\footnote{\url{http://archive.eso.org/bin/qc1_cgi?action=qc1_browse_instrume}. The quality control parameter {\tt model\_diffy\_med} for ThAr slit observations provides the difference to the model predictions.}.
We found that we have to adjust\footnote{We provide here the exact correction values up to the third decimal digits as we have applied them. This does not means that the precision of these correction is 0.001~\kms.} the UVB spectra by adding 4.477~km~s$^{-1}$, and the VIS spectra by adding 1.001~km~s$^{-1}$. 

After correction for these offsets, radial velocities (RV) measured from the UVB and VIS arms come in much better agreement (see also Sect.~\ref{ss:RV}). Figure~\ref{f:RV} is based on a preliminary determination of the radial velocities that processed the UVB and VIS arms separately. It
indeed shows that the systematic residual offset  is well below $1$~\kms, independent of the quality of the spectral lines of the object and of the \snr\ of the data. This amounts to about 10\%\ of the statistical uncertainties on these measurements.
The larger radial velocity differences for the lower-quality data are mainly due to emission lines, which are either stellar (in WR stars), or nebular.

\begin{figure*}
    \centering
    \includegraphics[width=17cm]{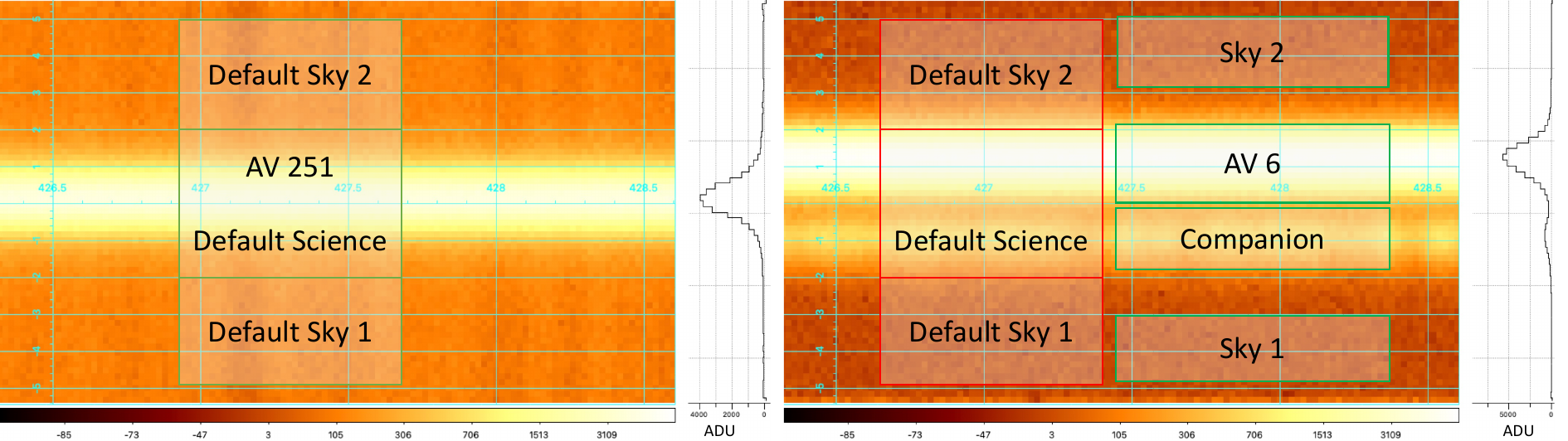}
    
    \caption{{  2D reduced spectrum of AV~251 (left) and AV~6 (right), revealing a faint companion 1\arcsec\ below the spectrum of AV6.  The overlaid grid indicates the wavelength axis (horizontal, in nm) and the spatial axis (vertical, in arcsec from the center of the slit). The default  science and sky extraction windows and those adopted for AV~6 in this work are overlaid on the 2D spectrum. The color bar indicate the ADU level in the images (in log scale).  The cross-sections along the spatial direction are in units of ADU (in linear scale). }}
    \label{f:sky}
\end{figure*}

\begin{figure*}
    \centering
    \includegraphics[trim={0 0 0 0.9cm},clip,width=\columnwidth]{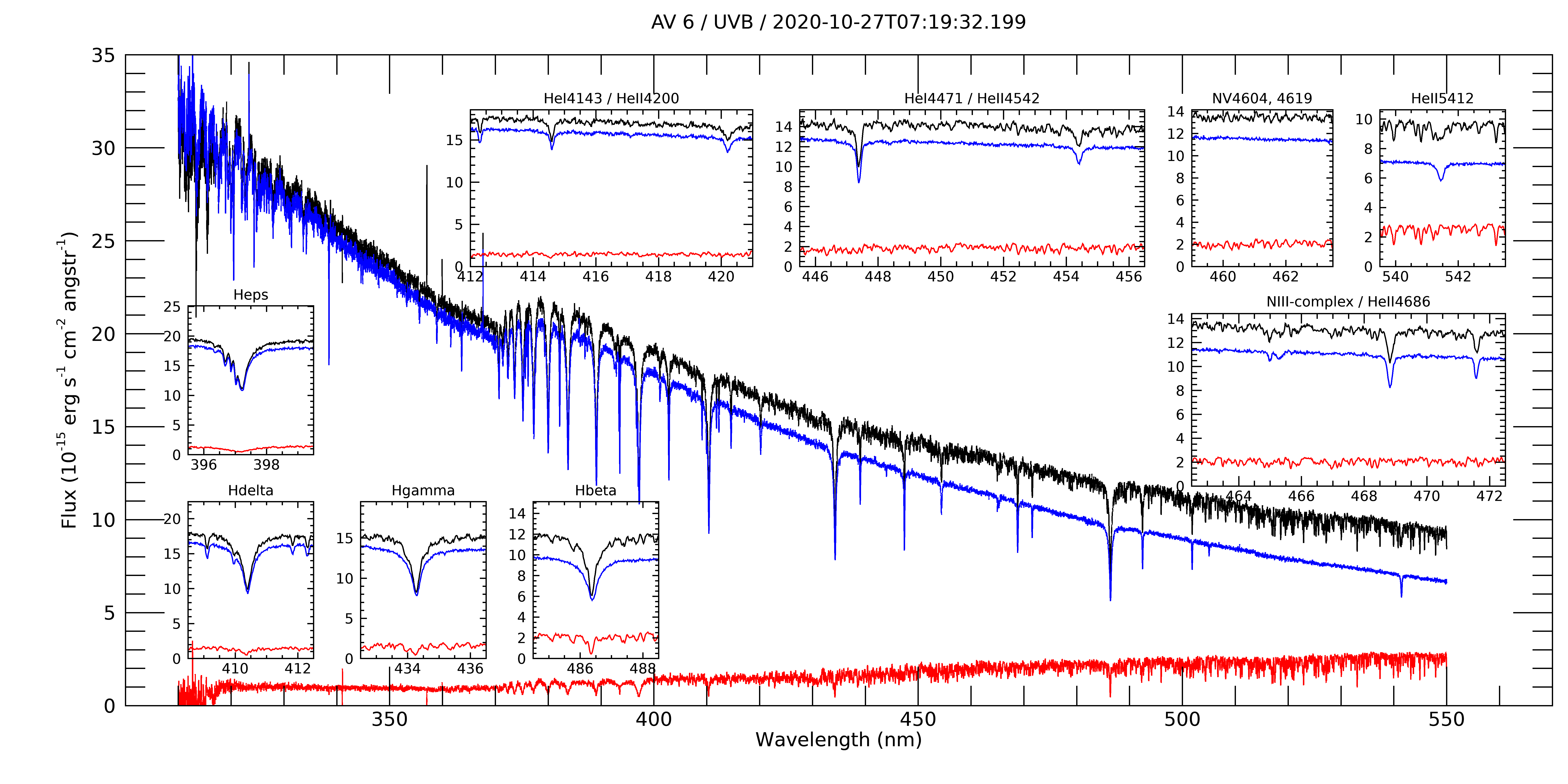}
    \includegraphics[trim={0 0 0 0.9cm},clip,width=\columnwidth]{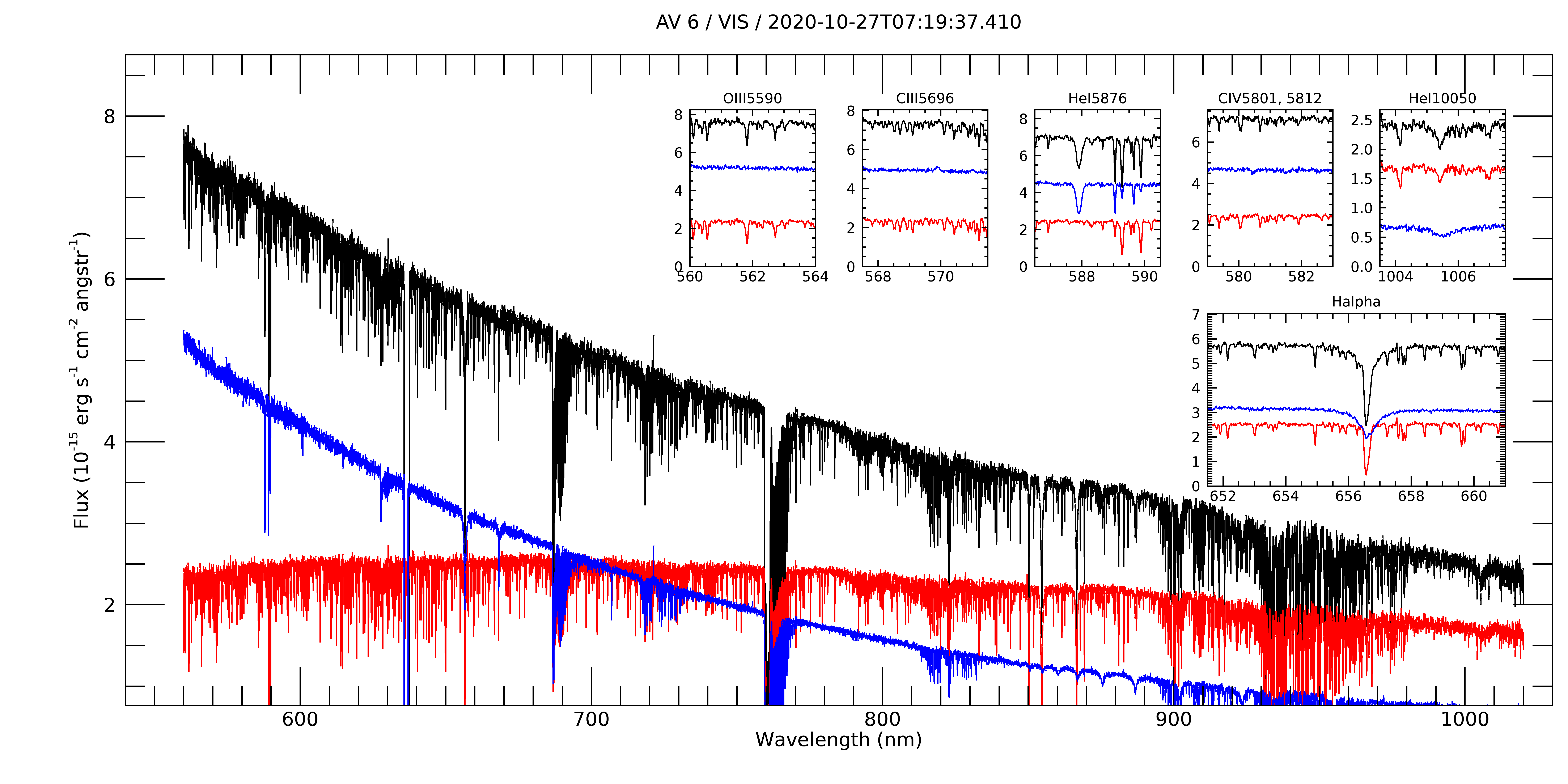}
    \caption{Comparison of the extracted spectra of AV~6 with the default (black) and manually selected extraction and sky regions for the components centered at +1.05\arcsec (blue) and $-0.8$\arcsec (red). Left- and right-panels correspond to the UVB and VIS arms, respectively. Inserts focus on various spectral lines of interests and have the same units as the main panel.}
    \label{f:AV6}
\end{figure*}

\subsection{Updated flux standard models / instrument response} \label{ss:flux} 

Following the flat fielding, the extracted spectra are corrected for the instrumental response curve  using a set of seven \xsh\ spectro-photometric white dwarf standard stars (GD~71, GD~153, EG~274, FEIGE~110, EG~21, LTT~3218, and LTT~7987) observed during the same or adjacent nights.  The standard reduction procedure in the public ESO pipeline \citep{Moehler2014} resulted in inconsistent spectral energy distribution slopes between the standard stars and small ``wiggles''  around the Balmer lines (on the order of a few percent, see Fig.~\ref{f:FluxModel}), which can affect the measurement of vital stellar parameters such as temperature and surface gravity. We updated the reference models and the knot anchors points used in the corresponding pipeline recipe to reduce these wiggles and boost the consistency in the relative flux calibration (Fig.~\ref{f:FluxModel}).  

Of the seven stars used as \xsh\ standards, two (GD~71 and GD~153) are also used as standards for the {\it HST\/}. For consistency with the flux calibrated {\it HST\/} spectra, we used the {\it HST\/} reference models for GD~71 and GD~153 as inputs for the \xsh\ pipeline \citep{Bohlin2020}.  To correct the models of the remaining five \xsh\ standards, we collected all their observed data between Oct 12, 2020 and Apr 21, 2021 and reduced them as if they were science frames, using the nearest GD~71 observations as the flux calibrator and the {\it HST\/} reference models for the reduction. In total, 110 such observations were available (4 for EG~274, 23 for FEIGE~110, 33 for EG~21, 38 for LTT~3218, and 12 for LTT~7987). We co-added all reduced spectra for each of these standards and performed an interpolated grid-based chi-square minimization to obtain the best-fitting synthetic white dwarf model for each.  

Given the parameter range covered by these five standard stars ($T_\textrm{eff} \approx [9, 45]$ kK, $\log(g) \approx [5, 8.5]$), the master grid used for the fitting consisted of models taken primarily from the WDGrid \citep{Bohlin2020}, supplemented by models from TMAPGrid2 \citep{Werner1999, Werner2003, Rauch2003}, and from a grid computed in \citet{Koester2010}.  After the best fitting model was determined, we followed the procedure outlined in \citet{Moehler2014} to remove the large scale bumps in the residuals that result from missing physics in the synthetic models using the SciPy UnivariateSpline method \citep{Virtanen2020}.  Finally, we adjusted the knot points used in the flux calibration pipeline so that they are more symmetric around the Balmer lines in order to reduce the wiggles initially seen in the reduction. Figures~\ref{f:FluxModel} and \ref{f:FluxModel_hbeta} illustrate the improved consistency obtained with the new flux standard models.

\begin{table}[]
    \caption{List of stars with multiple objects in the slit and adopted science and sky extraction regions, expressed in arcsec. The Sky 1 region is always located below the science spectrum on the detector while the  Sky 2 region is above (see Fig.~\ref{f:sky}).}    \label{t:sky}
    \centering
    \begin{tabular}{l c c c}
    \hline
    \hline
     Targets  &  Science & Sky 1 & Sky 2 \\
              & [start: end] &  [start: end] &  [start: end] \\
    \hline
\multicolumn{4}{c}{Multiple Objects on slit}\\
    \hline
AV~6        & $+$0.0:+2.1 & $-$3.0:$-$4.8 & $+$3.2:$+$5.1 \\
NGC346-ELS-026   & $-$2.0:+1.4 & $-$5.0:$-$2.0 & none\\

 N206-FS-170 & $-$2.0:+$2$.0 & $-$5.0:$-$2.0 & $+$4.0:$+$4.8\\ 
N11-ELS-048 & $-$2.0:$+$1.7 & $-$5.0:$-$2.0 & none\\
VFTS-482    & $-$2.0:$+$2.0 & $-$4.9:$-$3.7 & none\\
VFTS-542    & $-$2.0:$+$2.0 & $-$5.0:$-$2.0 & none\\
VFTS-545    & $-$2.0:$+$2.0 & $-$5.0:$-$2.0 & none\\
    \hline
\multicolumn{4}{c}{Non-standard extraction of isolated stars}\\
    \hline

AV~324   & $-$3.0:$+$3.0 & $-$5.0:$-$3.0 & $+$3.0:$+$5.0 \\ 
SK-67D106&  $-$3.0:$+$3.0 & $-$5.0:$-$3.0 & $+$3.0:$+$5.0 \\
SK-191   &   $-$3.0:$+$3.0 & $-$5.0:$-$3.0 & $+$3.0:$+$5.0 \\
    \hline
    \end{tabular}
\end{table}

\subsection{Sky subtraction} \label{ss:sky} 
All advances products that we provide are sky subtracted, that is the average flux level of an {\it empty} region of the detector has been subtracted from the flux of the science object. According to this definition, the {\it sky signal} contains the illumination of the night sky (moon, dark sky brightness) as well as any nebulosity or extended emission that will fall on the slit. We effectively work under the assumption that these latter components are not variable. While all the advanced products provided are sky-subtracted, we provide the sky spectrum used so that it is trivial for any user to reconstruct the original, non-sky subtracted  signal. We briefly describe below the sky subtraction algorithm as well as the procedure applied when multiple objects  where visible in the 2D non-sky subtracted spectra.

The sky level is estimated from regions outside the object mask  and away from the detector edges by more than 0\farcs5 (approx.~3 pix). 
The object mask extends over $\pm$2\arcsec, thus $\approx$25-pixels-wide around the centre of the slit. This leaves two regions of approx.\ 3\arcsec\ above and below the science spectrum where the sky was estimated from.  The selected sky pixels are then tabulated as a function of wavelength and the 2D spectrum is then smoothed along the wavelength axis using a running median over 51 pixels.

For ten targets, one or more additional objects are present in the slit beside the main target. These contaminants were identified by visual inspection of the \raw\ images and their cross-dispersion profile. For these objects, the sky regions were selected manually upon visual inspection of the reduced 2D spectra. The adopted sky regions are given in Table~\ref{t:sky}. An example is shown in Figs.~\ref{f:sky} and \ref{f:AV6}. Additional comments are provided in App.~\ref{app:targets}.

\subsection{Cosmic ray hits} \label{ss:cosmic} 
\begin{figure}
    \centering
    \includegraphics[width=\columnwidth]{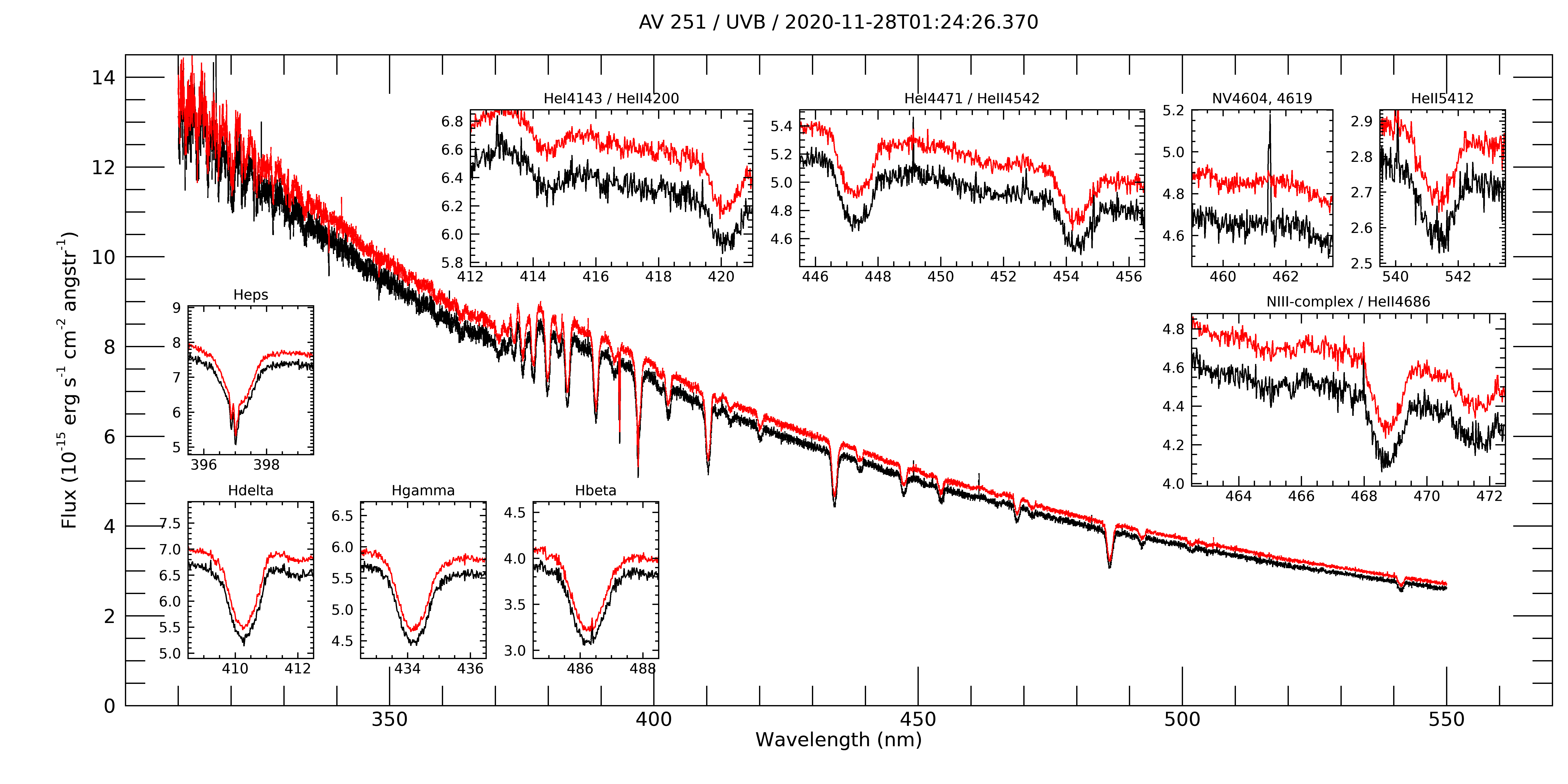}
    \caption{Comparison of reduced spectra of AV~251 obtained with the different cosmic rejection thresholds. Black: {\tt -valueremovecrhsingle-sigmalim=20} (default). Red: {\tt -valueremovecrhsingle-sigmalim=10} (adopted). Spectra are slightly shifted for clarity. Inserts focus on spectral lines of interest and have the same units as the main panel.}
    \label{f:CR}
\end{figure}

The ESO CPL pipeline allows to flag cosmic ray hits by  using a Laplacian edge detection algorithm \citep{2001PASP..113.1420V}. The default detection threshold ({\tt -valueremovecrhsingle-sigmalim=20}) resulted in a significant number of cosmic ray hits retained in the extracted 1D science spectra. Experimenting with different values, we opted for  a  detection threshold at $10\sigma$ ({\tt -valueremovecrhsingle-sigmalim=10}). Figure~\ref{f:CR} shows a comparison of the extracted 1D spectra with the default and with our adopted value.

\subsection{Order re-connection} \label{ss:order}
Order re-connection in \'echelle spectrographs are a frequent issue and we paid particular attention to check the quality of the extracted and merged 1D science spectra. Poor order re-connection does not only complicate the normalisation process, but may induce significant biases in the science analysis if the impact on the shape or strength of spectral lines is unidentified. In the \xshooter\ pipeline, order reconnection is implemented as  a weighted mean in the overlapping parts of the orders. The weight are defined as the inverse of the variance, hence $1/\sigma^2$ where $\sigma$ is the signal-to-noise ratio. This approach is indeed statistically correct in case of a normally distributed noise and no systematic biases.

\begin{figure}
    \centering
    \includegraphics[trim={0 0 0 0.9cm},clip,width=\columnwidth]{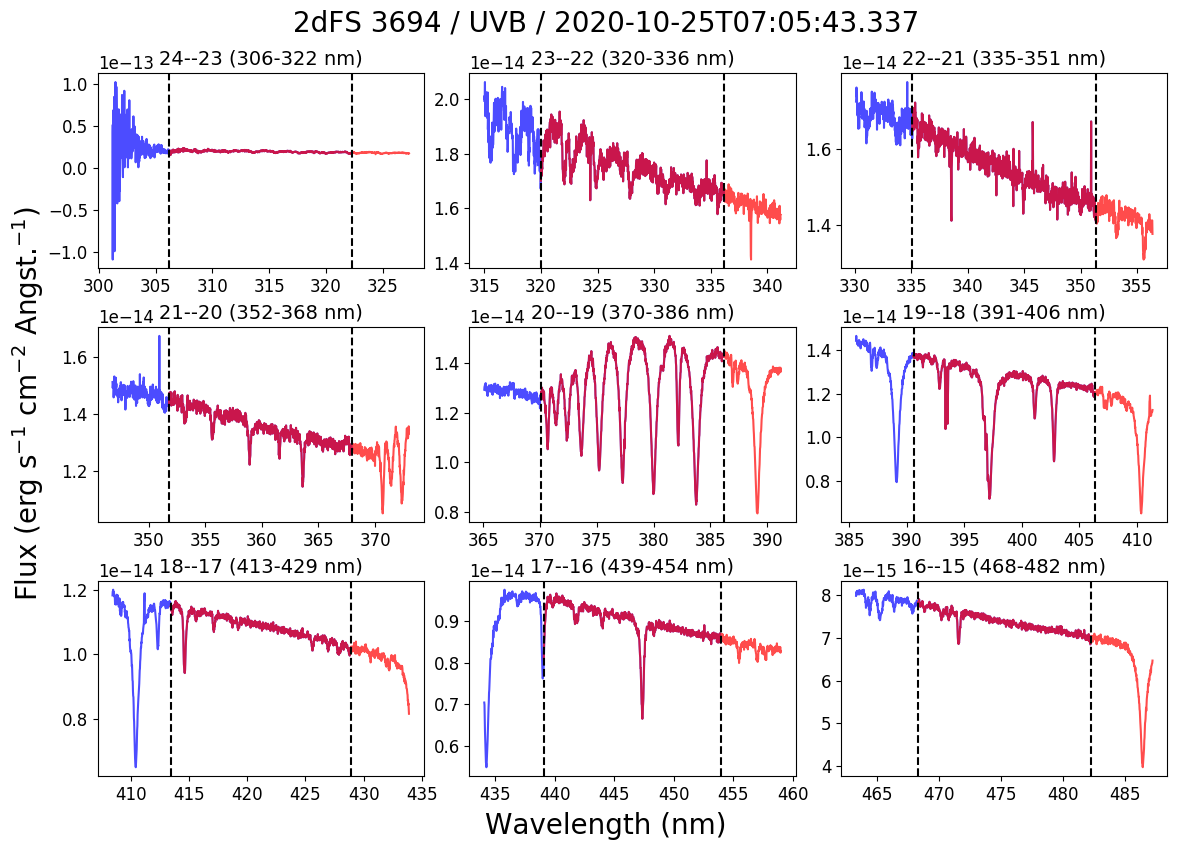}
    \caption{Examples from the reduced UVB spectrum of 2DFS-3694 illustrating the absence of order re-connection artefacts. Blue and orange lines correspond to the wavelength regions of the first and second orders (as presented at each sub-panel). Vertical dashed lines indicate the locations of the order re-connections (highlighted as red lines, which is the overlapping of the blue and orange regions). Order numbers and the common wavelength range of the orders are indicated on top of each sub-panel. }
    \label{f:order}
\end{figure}

In this context, we visually inspected each re-connection region (see Fig.~\ref{f:order} for an example) and concluded that the order re-connection in the response-corrected  spectra was excellent and void of any systematic. 

\section{Advanced data products}\label{s:ADP}

\subsection{Slitloss correction} \label{ss:slitloss}  

The XShootU spectra were obtained with narrow slits (0\farcs8 and 0\farcs7 for the UVB and VIS arms, respectively). The width of the entrance slits is thus of the same order of magnitude as the median seeing value at Cerro Paranal \citep{2008Msngr.132...11S} and thus a fraction of the star PSF does not enter the spectrograph slits. The \flux\ standard star observations used to determine the response curve correction (Sect.~\ref{ss:flux}) were observed with 5\arcsec-wide slits such that the entire star PSF is transmitted through the slit. To obtain absolute flux calibrated spectra, we corrected the response-curve corrected spectra for slit losses due to seeing and image quality across the detector. 
The correction for  slit-losses is wavelength dependent, because the seeing value ($\sigma_{\lambda}$) changes with wavelength.

\begin{figure*}
    \centering
   \includegraphics[width=17cm]{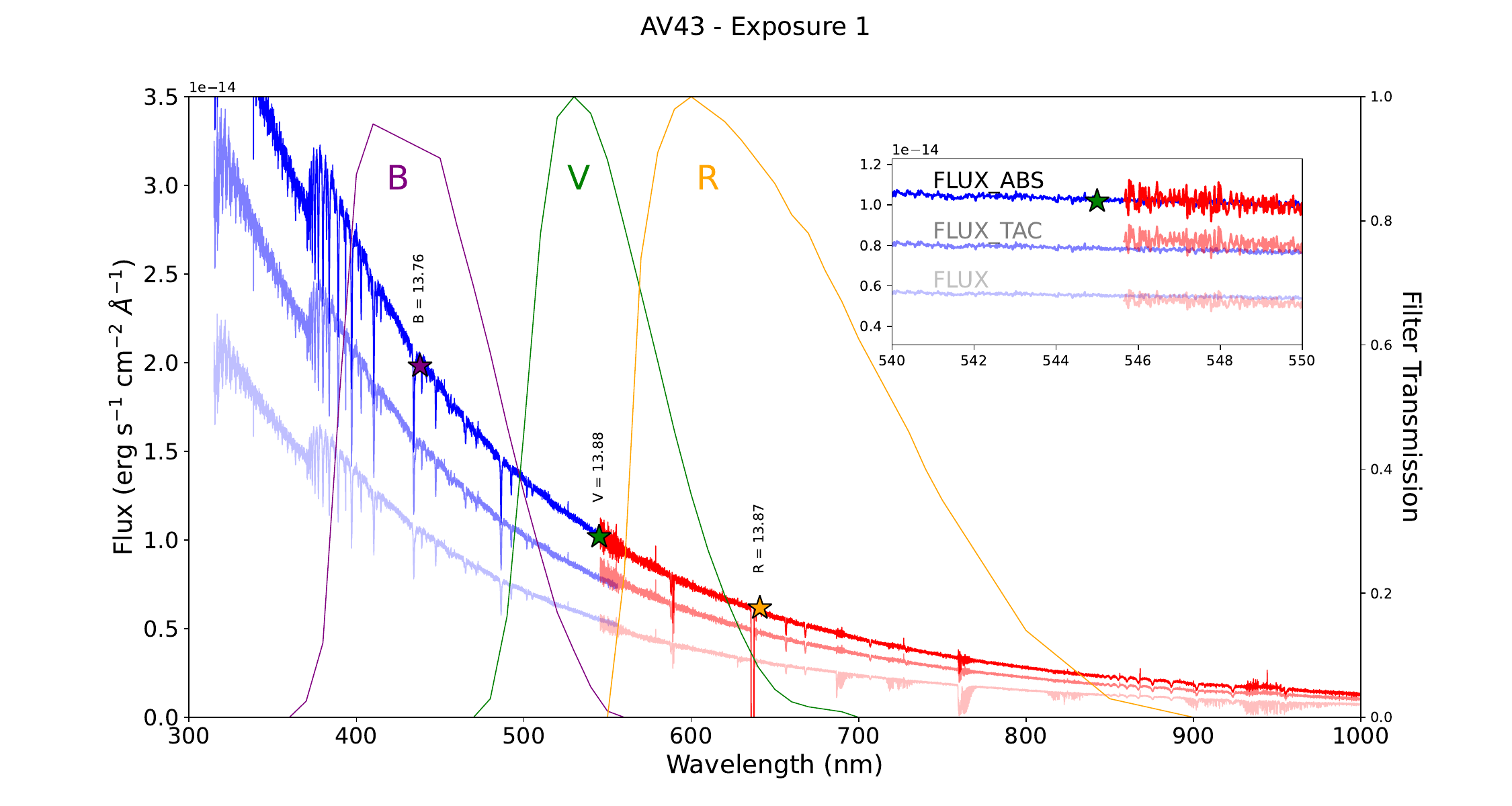}
    \caption{Comparison of the spectrum of AV-43 before slit-loss correction (lower spectrum), after slit-loss and telluric absorption correction (middle spectrum), and after the scaling to the B magnitude and lining up the VIS to the UVB (upper spectrum). The corresponding V and R magnitudes are indicated using a star symbol. Transmission curves are from \citet{bessel90}.}   \label{f:FluxCal}

\end{figure*}

The seeing in Paranal is measured in real-time by a differential image motion monitor (DIMM). Using the DIMM values recorded in the ESO archive, we estimated the average seeing  during each \xshootu\ observation. As the DIMM seeing ($\sigma_\mathrm{DIMM}$) is defined at 500~nm at zenith, this value needs to be corrected for the wavelength dependence and the airmass of each observation \citep{1966JOSA...56.1372F, 2000stop.book.....G}:
\begin{equation}
    \sigma_{\lambda} = \sigma_\mathrm{DIMM} \left({\frac{\lambda}{500~\mathrm{~nm}}}\right)^{-0.2} \times \mathrm{airmass}^{0.6}.
\end{equation}
The slit transmission fraction ($T$) at each wavelength is given by:
\begin{equation}
    T = \mathrm{erf} \left(\sqrt{\ln{2}}\hspace*{1mm} \frac{w_{\mathrm{slit}}}{ \sigma_{\lambda}}\right),
\end{equation}
where $w_{\mathrm{slit}}$ is the slit width and $\mathrm{erf}(x)$ is the error function:
\begin{equation}
    \mathrm{erf}(x) = \frac{2}{\sqrt{\pi}} \int_0^x e^{-t^2}dt.
\end{equation}
We then corrected the (relative) response-corrected spectrum for the slit losses ($1-T$) at each wavelength bin. An example is given in Fig.~\ref{f:FluxCal}.

\subsection{Telluric Correction / Molecfit} \label{ss:telluric}

\begin{figure*}
    \centering
    \includegraphics[width=\textwidth]{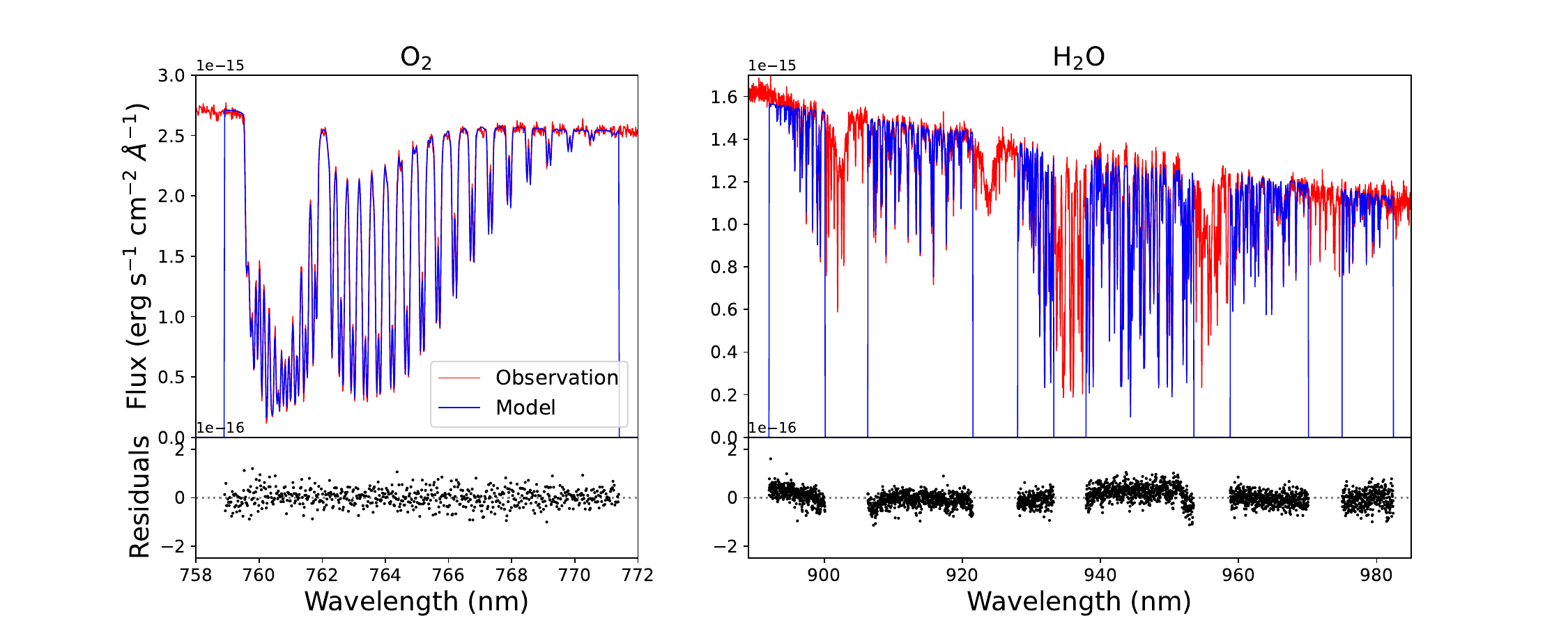}
    \caption{Example of a telluric model fit to the strong O$_2$ (top left) and H$_2$O (top right) bands for one of the exposures of AV-43. The residuals of the fit (observation - model in $10^{-16}$\,erg\,s$^{-1}$\,cm$^{-2}$\,\AA$^{-1}$) are shown in the bottom panels.}
    \label{f:QCtellfit}
\end{figure*}

\begin{figure*}
    \centering
    \includegraphics[width=17cm]{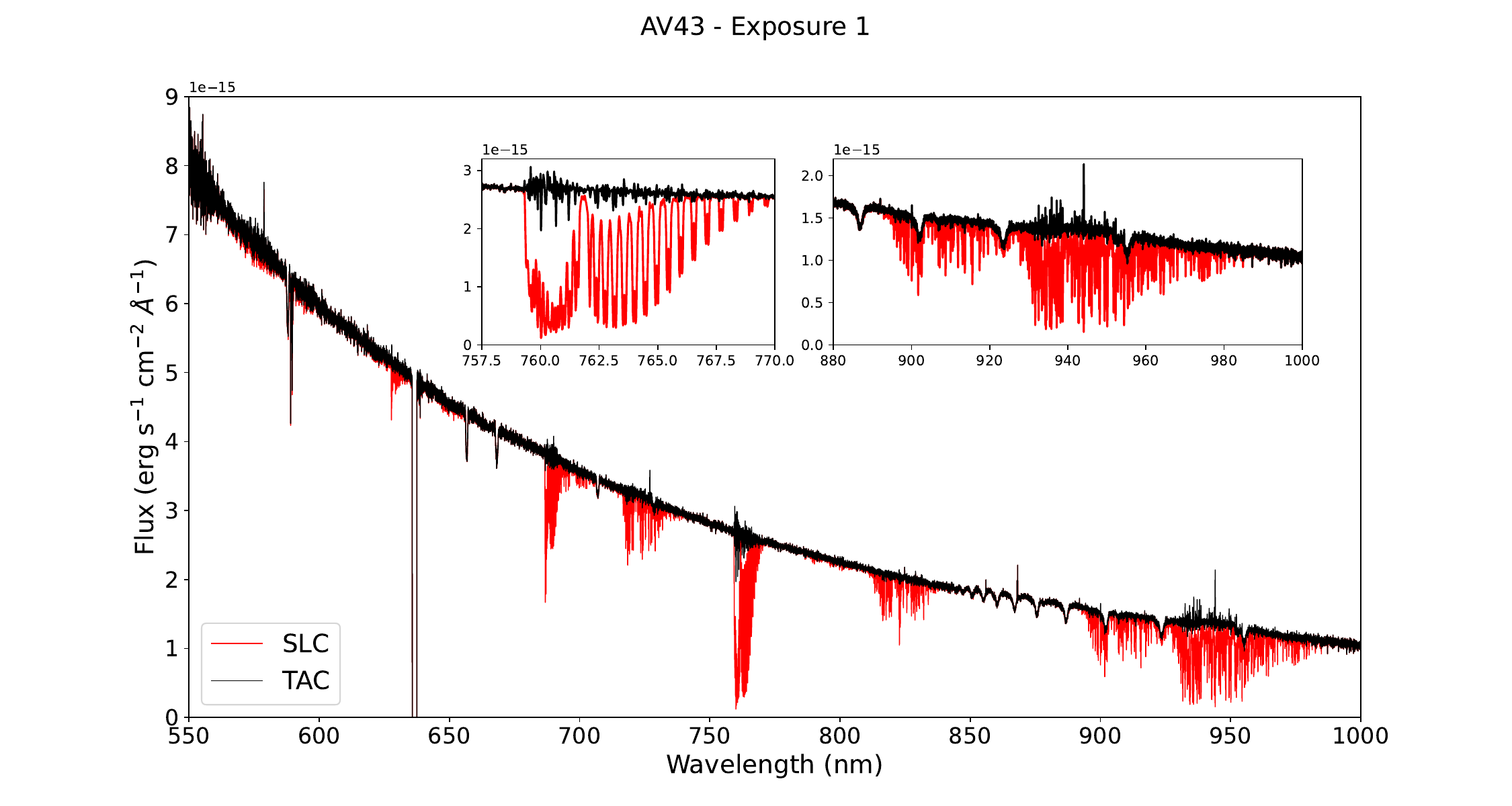}
    \caption{Comparison of the spectrum of AV-43 before (SLC) and after (TAC) using the model from Fig.~\ref{f:QCtellfit} to correct for telluric absorption. The insets show the two regions with the strongest absorption features; they have the same units as the main panel.}
    \label{f:QCtell}
\end{figure*}

The slit-loss corrected spectra of the VIS arm were corrected for telluric absorption using the {\sc molecfit} tool v3.0.3 \citep{2015A&A...576A..77S,2015A&A...576A..78K}. The telluric model includes O$_2$ and H$_2$O absorption lines, appropriate for the VIS wavelength region. The standard wavelength regions used by {\sc molecfit} were adjusted to provide the best overall results for the full sample of stars. In particular, all wavelength regions with strong telluric features are included, while regions with stellar spectral lines are excluded. The adopted wavelength ranges are given in Table~\ref{t:TAC}. 

Generally, the spectrum can be well fitted by the telluric model, with typical fit residuals on the order of 1-2\%\ (Fig.~\ref{f:QCtellfit}). This results in a good telluric correction in most of the wavelength regions with weak telluric lines, with residuals well below 1\%\ (see Figs.~\ref{f:QCtell}). Notable exceptions are the strongest molecular bands of O$_2$ at 760-770~nm and H$_2$O at 930-980~nm, both of which are close to or fully saturated (depending on weather conditions). Residuals in these regions typically reach up to 10\%\ (see Fig.~\ref{f:QCtell}, insets), but can be much higher in rare cases with particularly poor weather conditions.

\begin{table}[]
    \caption{Wavelength ranges used for fitting the telluric spectrum (in nm).}
    \centering
    \begin{tabular}{c|c}
    \hline
    \hline
     Included  & Excluded \\
    \hline
    584.1 - 603.0 & 587.0 - 591.0 \\
    627.4 - 632.4 & 821.5 - 826.0 \\
    640.4 - 650.0 & 900.1 - 906.3 \\
    685.0 - 697.0 & 921.5 - 928.0 \\
    697.1 - 705.0 & 933.2 - 937.9 \\
    722.7 - 726.9 & 953.5 - 958.8 \\
    758.9 - 771.4 & 970.1 - 975.0 \\
    810.1 - 821.0 & \\
    821.1 - 834.4 & \\
    892.0 - 982.4 & \\
    
    \hline
    \end{tabular}
    \label{t:TAC}
\end{table}

\subsection{Absolute flux calibration}\label{ss:absfluxcal}

After the  response-curve and slitloss correction steps (Sects.~\ref{ss:flux} and \ref{ss:slitloss}), one might expect that the relative scaling between the UVB and VIS arms are correct and that the integration of absolute fluxes over, e.g., the $B$ or $R$-bands, matches the corresponding observed photometric magnitudes. However, despite considerable improvements, the agreement is generally not perfect and differences of about 5\%, on average, between both arms and with respect to published photometry remain. These likely arise from residuals from arm-dependent flux losses, 
sky subtraction, and variations in the sky transparency, seeing, and airmass during the integration time. It might also result from differences between the DIMM seeing and the true seeing.  We thus implemented
 two additional steps to further improve on the absolute flux calibration of the \xshootu\ spectra. 

\begin{figure}
 \centering
    \includegraphics[width=\columnwidth]{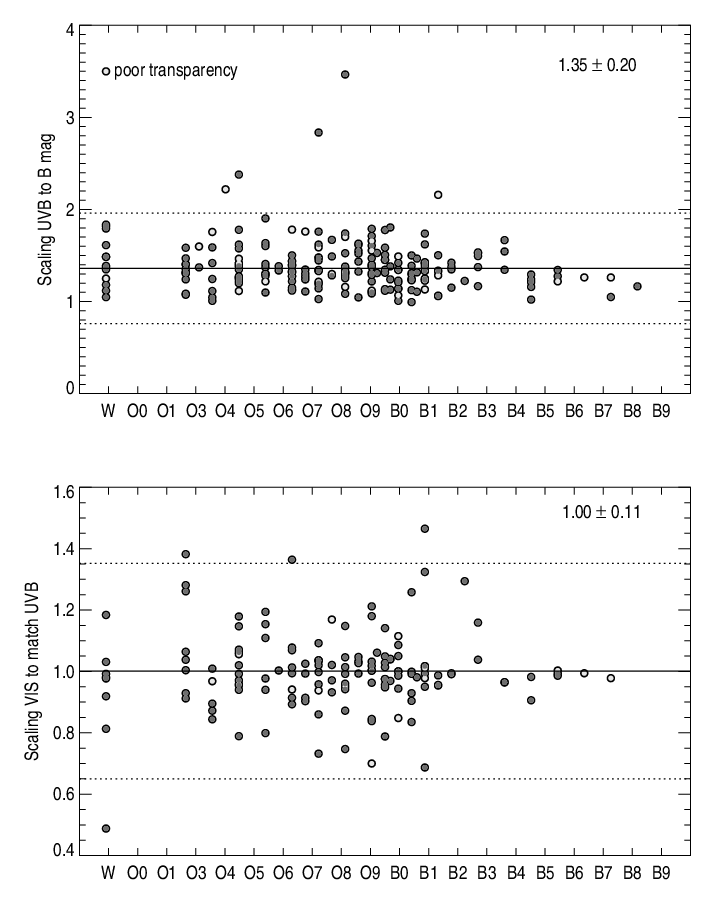}
     \caption{Scaling factors for absolute flux correction, as a function of spectral type. Top panel: scaling factors applied to the UVB arm to make the spectra match the B magnitude photometry. Bottom panel: extra scaling factors applied to the VIS arm, after matching them to the R photometry, to align the UVB and VIS arms. Dotted lines in both panels correspond to $\pm3\sigma$ ranges.}
    \label{f:scallfact}
\end{figure}

In a first step, we scaled the slit-loss corrected spectra to published photometric measurements of our targets. To do this, we collected $B$ and $R$ magnitudes for all our targets. All used $B$ magnitudes and their references are listed in Table~1 of \citetalias{xsh1}, with the exception of the magnitudes for AV232, HV5622, LH58-496, NGC346-ELS-25, NGC346-ELS-51, VFTS-169, and VFTS586, which come from \citet{massey2002}. All $R$-band photometry comes from \citet{massey2002}, with the exception of AV264, which comes from \citet{zacharias2013}.
The scaling factor was calculated using the $B$ filter for the UVB arm and the $R$ filter for the VIS arm as these filter's bandpasses are almost entirely contained within the wavelength range of the two arms (see Fig.~\ref{f:FluxCal}). Transmission curves were adopted from \citet{bessel90}. $U$- and $V$-band values were used to check the consistency of the applied scaling. 
Figure~\ref{f:scallfact}, top panel, shows the distribution of scaling factors for the UVB arm, per spectral type. 

After this step, small shifts between the UVB and VIS arms remained for many observations (3\%\ on average), e.g., due to uncertainties in the published photometry or target variability. 
To create a smooth SED across both arms, we applied a second scaling. We calculated the average flux in each arm from 545~nm to 555~nm (the overlapping wavelength region), and applied a global  scaling factor to the VIS arm to align it with the flux of the UVB arm. For most observations, this correction was less than 10\%. For targets with no published photometry, we only applied this second step. For some targets, no $R$-band magnitudes are available. In this case, we scaled only the UVB spectra to published photometry and the second scaling step was used to align the UVB and VIS arms. 
Figure~\ref{f:scallfact}, bottom panel, shows the distribution of the additional scaling factors for the VIS arm, per spectral type. This figure  only includes the stars for which $R$-band photometry was available. For both panels,  no trend with spectral type is visible, and variations must be connected to  seeing variations during observations, variable cloud coverage or the uncertainties in the photometry values.

A comparison of spectra before and after slit-loss correction, and after the final scaling to the $B$ magnitude, is shown in Figure~\ref{f:FluxCal}.

\begin{figure*}[t!]
    \centering
    \includegraphics[width=\columnwidth]{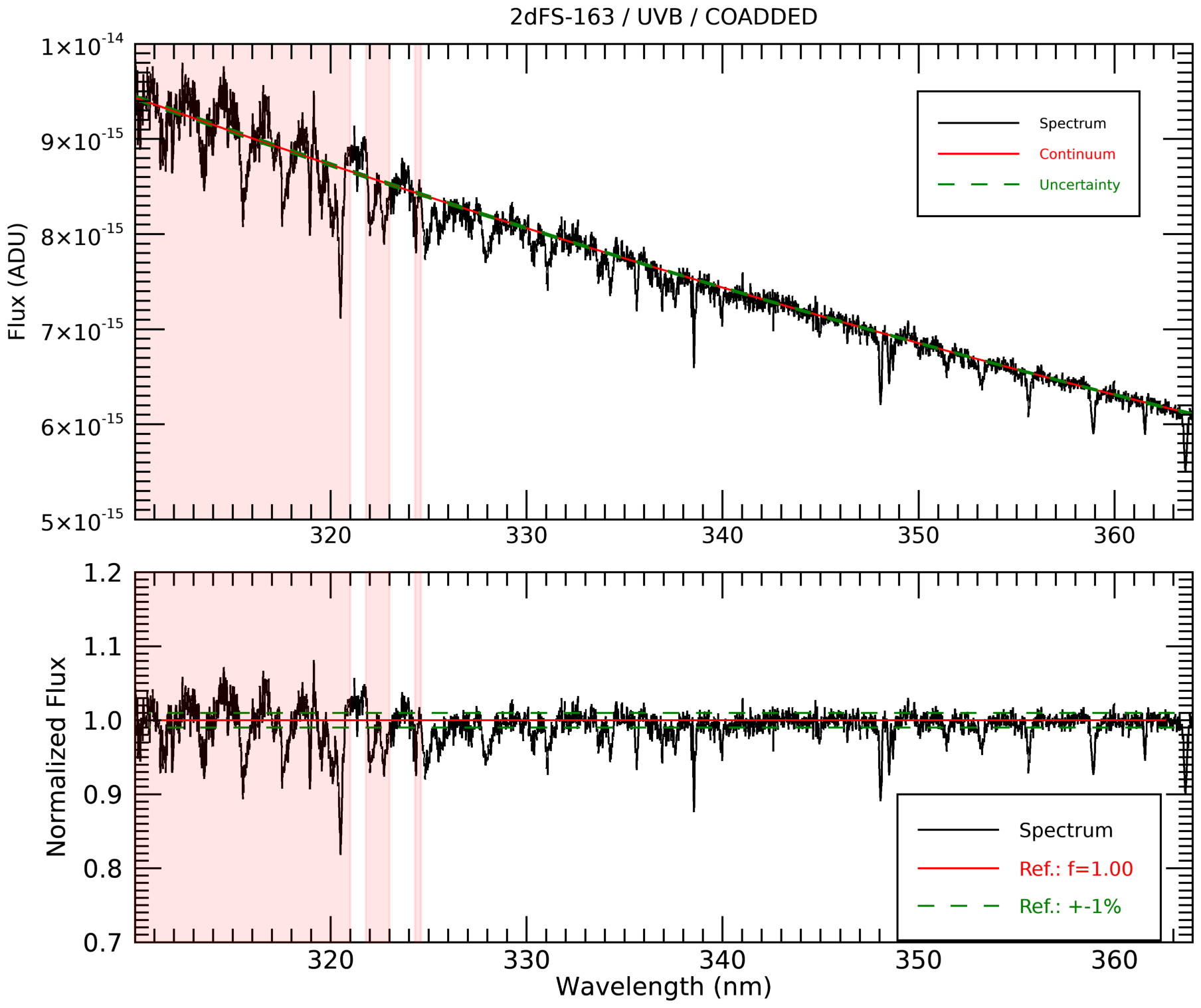}
    \includegraphics[width=\columnwidth]{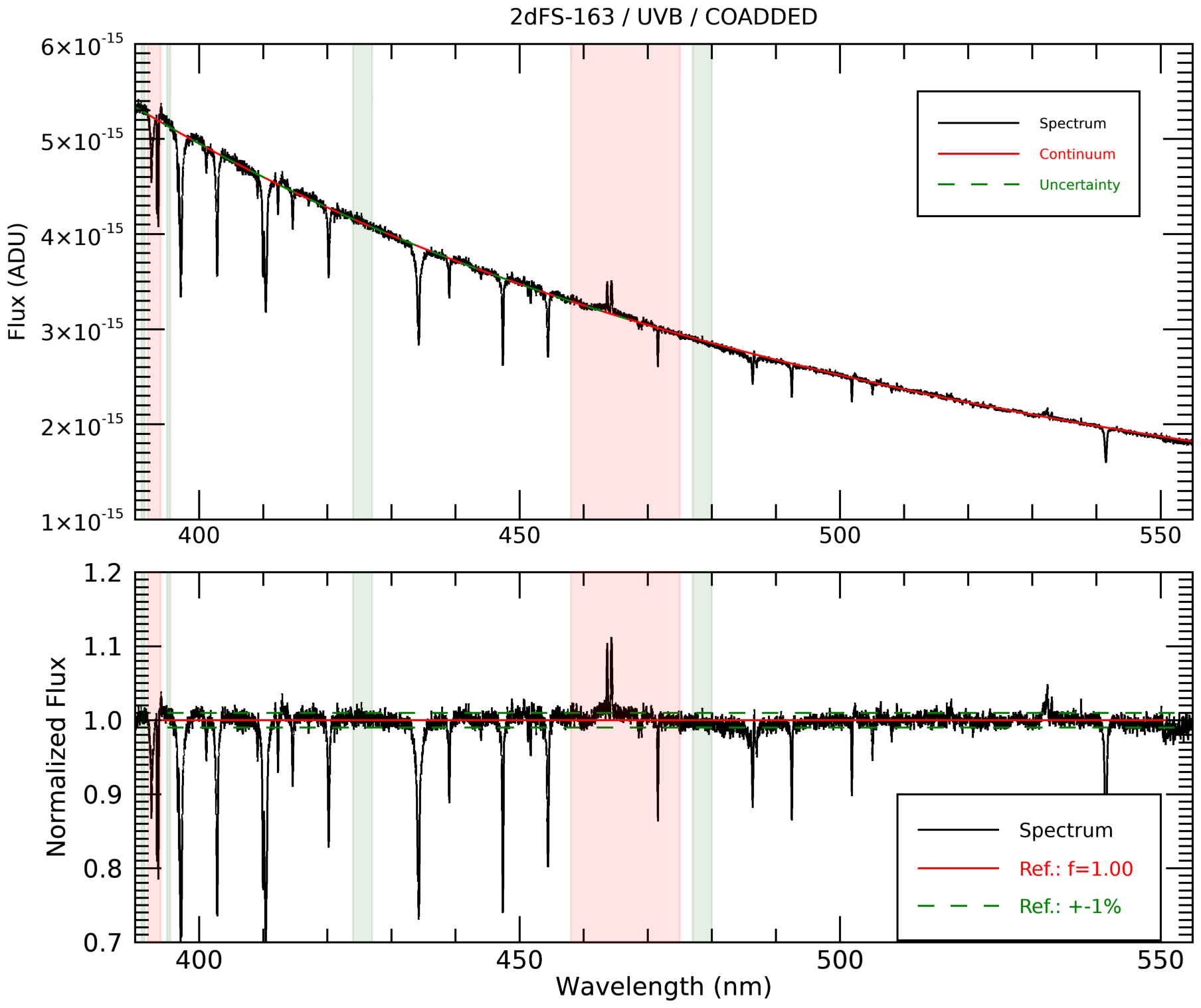}
    \includegraphics[width=\columnwidth]{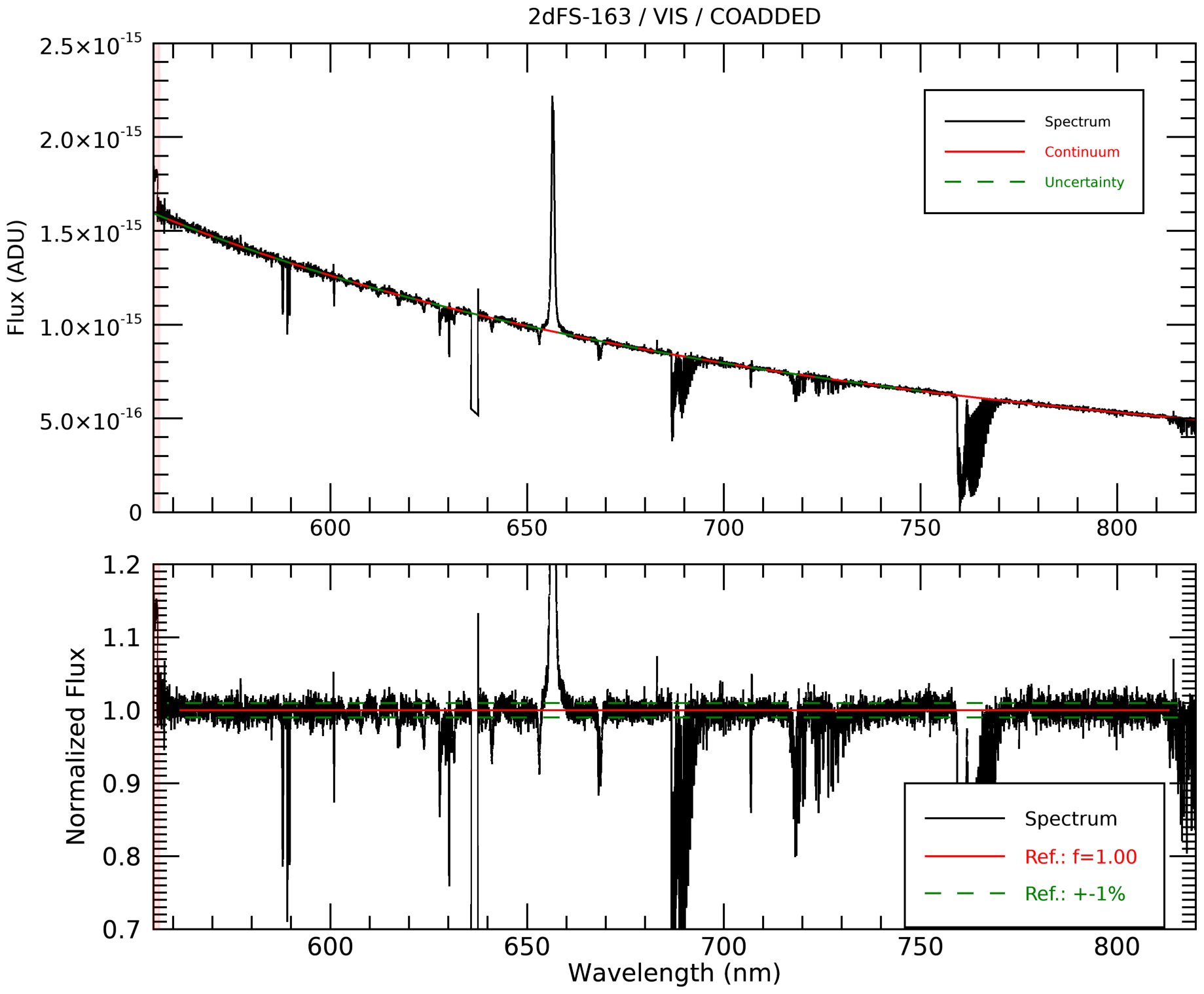}
    \includegraphics[width=\columnwidth]{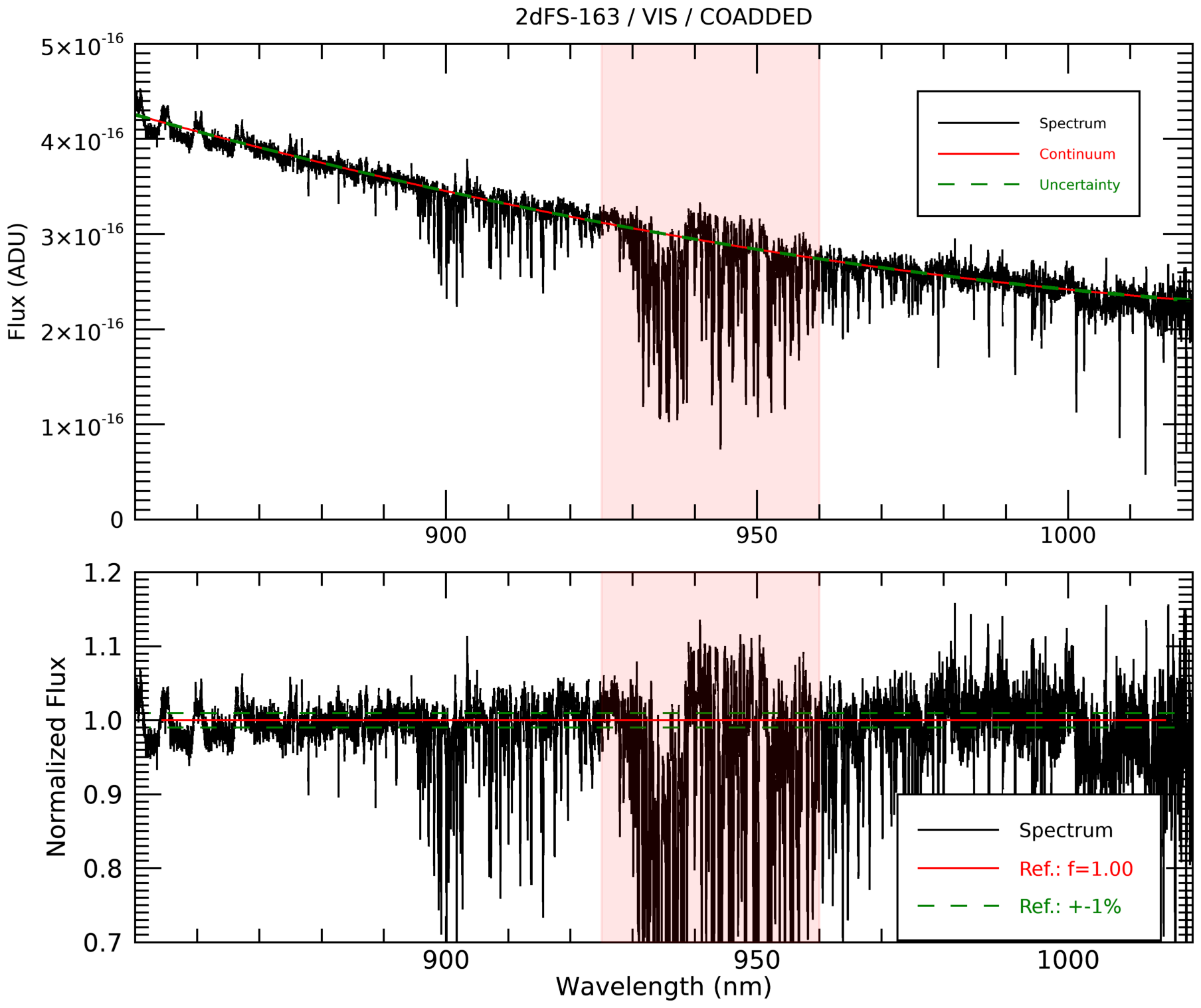}
    \caption{Example of automatic normalisation of 2DFS-163 for each range in the UVB and VIS arms. Regions shaded in red and green are regions that are rejected or forcefully included as continuum regions. }
    \label{f:QCnorm}
\end{figure*}

\subsection{Rectification to continuum}  \label{ss:norm}
The  response-curve and slit-loss corrected individual exposures were individually normalised by adjusting the continuum with a functional form. The rectification procedure follows the general algorithm described in \citet{vfts8} but uses different functional forms. In short, the algorithm works in a pseudo wavelength range (centered on 0 and ranging from $-1$ to $+1$ for numerical stability) and proceeds in three steps: (i) an initial detrending of the flux calibrated spectrum using a linear fit; (ii) the identification and removal of absorption and emission line regions  using the ratios of a min/max filter with a median filter, and (iii) an iterative $\kappa-\sigma$ clipping to adjust a pre-defined function to the remaining continuum points. Three iterations were usually adopted, but this could be modified upon a visual inspection of the results. 

To make the automatic rectification process more robust, we selected regions where no continuum points were allowed (e.g., regions around \heb~\l4686 or \halpha) but also anchor continuum regions where the $\sigma$-clipping could not remove continuum points (see Fig.~\ref{f:QCnorm}). These anchor regions help the rectification process to more robustly converge in case the first step in the iterative cleaning of spectral lines is too aggressive.

The full spectral range was divided in four  regions,  two in each arm, that were separately rectified (see Fig.~\ref{f:QCnorm}). For the UVB arm, the region blue-ward of the Balmer jump ($310-365$~nm) was rectified using a quadratic function while the region red-ward of the Balmer jump ($390-555$~nm) uses a generalised Planck-function:
\begin{equation} \label{eq:planck}
f_\mathrm{c}(\lambda) = a \lambda^b \left(e^{c/\lambda} -1 \right)^{-1},
\end{equation}
with $a$, $b$ and $c$ being variables of the fit.
Similarly, the VIS arm wavelength  range was split into two sections. The range blue-ward of  the Paschen jump ($555-820$~nm) was adjusted with Eq.~\ref{eq:planck} while the range redward of the Paschen jump ($800-1020$~nm) was adjusted with a quadratic function. 

The choice of the functions was motivated by  the respective sizes of the wavelength ranges to fit as well as by our intention to avoid introducing high-order frequencies in the rectified signal. Such high-frequency fluctuations could impact shallow absorption or emission profiles that are of astrophysical significance (e.g., extended Balmer wing absorption, diffuse interstellar bands or extended wind line emission) that would be over-corrected using a local approach or a more aggressive normalization function (e.g., high order polynomials, splines). In doing so, we provide a global high-quality normalisation (within 0.5\%\ of the continuum in most of the spectral range). Further local normalisation may need to be applied, depending on the science case.

The regions around the Balmer and Paschen jumps were excluded from the rectification procedure because these cannot be adjusted without an atmospheric model. As described above, the overall result is quite satisfying given the large spectral ranges covered (up to 265~nm) and the limited number of free parameters (only 3) in each region. Some limitations are, however, apparent: 
\begin{enumerate}
\item[-] The bluest region ($<350$~nm) displays only a pseudo continuum due to the high density of spectral lines. This prevented us to retain continuum points so that the adopted continuum is an interpolation of the quadratic function used in the $324.6-364.0$~nm region. 
\item[-] There is a tendency to underestimate the continuum below 410 nm, while overestimating the continuum just redwards of \hdelta\ (by $\sim$0.5\%); see Fig.~\ref{f:QCnorm}, top-right panel.
\item[-] Small ($\sim$0.5\%) dips in continuum bluewards of the Balmer lines are often present.
\item[-] Broad bumps ($\sim$1\%) in flux around 570 nm caused by the O$_2$ model of the telluric absorption correction (see Section~\ref{ss:telluric}).
\item[-] Broad bump in flux from $900-960$~nm. This is a result of the telluric absorption correction as the telluric lines are (close to) saturated in this region.
\end{enumerate}

\subsection{Co-addition and final products}\label{ss:coadd}  

The 1D flux-calibrated spectra corresponding to individual exposures of each arm were corrected for barycentric motion. They were also re-sampled to a common wavelength grid starting at 300~nm and 545~nm for the UVB and VIS arm, respectively, adopting a constant wavelength step of 0.02~nm (see Sect.~\ref{s:DR}). The re-binned spectra of all exposures were co-added using a weighted mean where the weight is given by the inverse of the square error spectrum provided by the pipeline.  Regions of 1D spectra that were identified as saturated (Sect.~\ref{s:QC}) were masked in the co-addition process.

We do not co-add the normalised spectra of each epoch but re-do the normalisation after co-addition of the flux-calibrated  UVB and VIS spectra using the same procedures as described in Sect.~\ref{ss:norm}. This avoids that any uncertainty on the rectification of the individual epochs is propagated to the co-added data.

Finally, the co-added flux-calibrated  spectra of the UVB and VIS arms were merged by stitching them together at $\lambda=550$~nm. We choose to adopt a strict transition from one arm to another rather than, e.g., a progressive transition over a small wavelength range. This choice was driven by the fact that each arm has a different spectral resolving power, such that it is clear for the user at which wavelength the change of resolution occurs. Data above 550~nm for the UVB arm and below 550~nm for the VIS arm are still made available in the data products of the individual epochs (see Sect.~\ref{ss:1D-DP}), but are ignored in the coadded data products (Sect.~\ref{ss:coadd}).

\subsection{Overall quality: S/N distribution}\label{ss:snr}  
\begin{figure}
    \centering
    \includegraphics[width=\columnwidth]{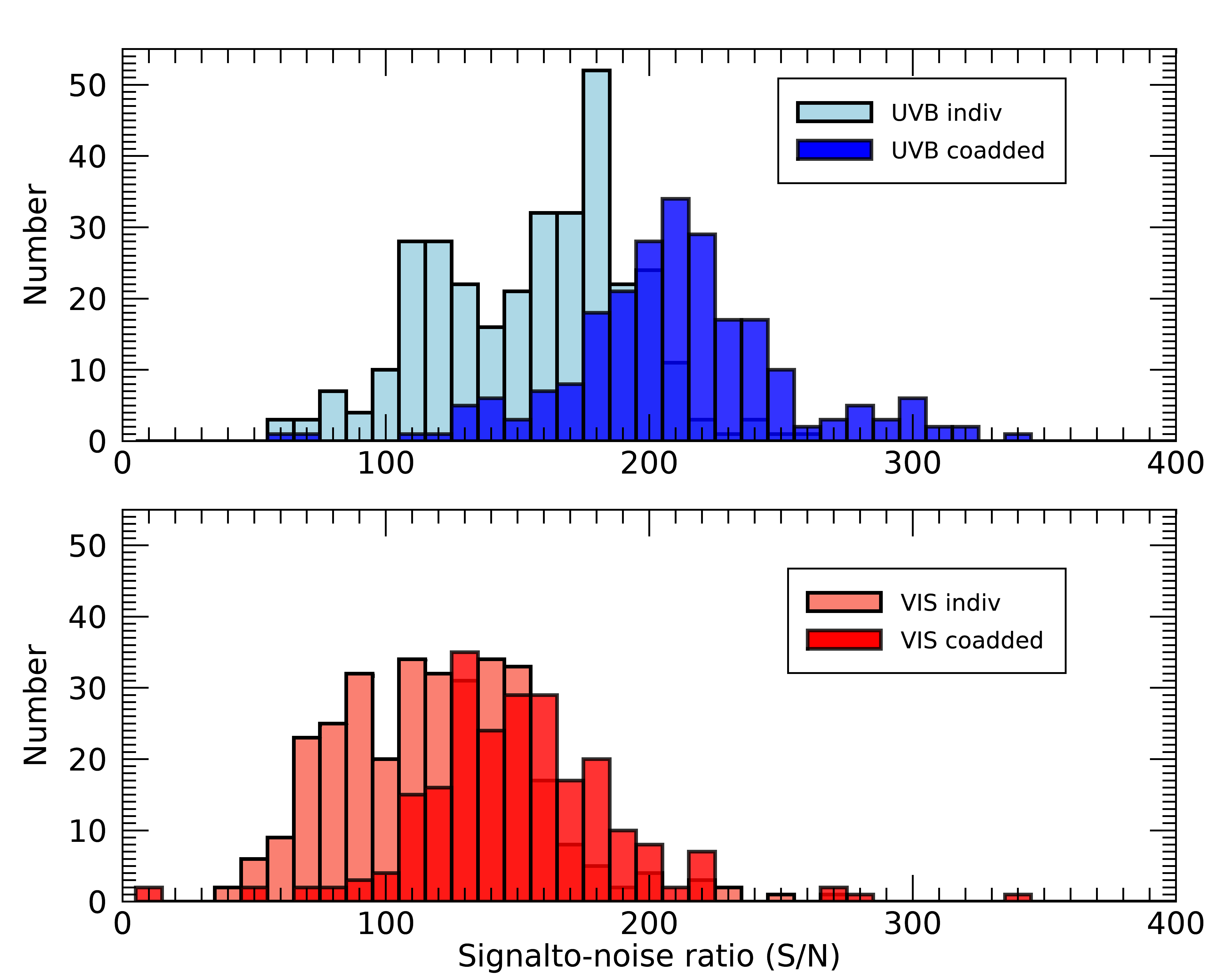}
    \caption{Distribution of \snr\ for the UVB (upper panel) and VIS (lower panel) part of the spectra.}
    \label{f:snr}
\end{figure}

We estimated the \snr\ obtained for each observation using continuum windows in the reduced and normalised spectra. For the UVB arm, the continuum region was taken from 475 to 480~nm while it ranged from 675 to 685~nm in the VIS arm. An exception occurs for broad emission line stars for which no automatic normalisation was applied. In these cases, the VIS continuum region was restricted to the range of 683 to 685~nm and a local normalisation was applied before computing the \snr. In each case, the \snr\ was computed per pixel as the ratio of the median to the 1$\sigma$-dispersion. The  resulting distributions are displayed in Fig.~\ref{f:snr} and show a median (per pixel) \snr\ above 200 in the UVB arm and around 150 in the VIS arm of the co-added spectra. These correspond to a median \snr\ per resolution elements of above 350 and 250 in the UVB and VIS coadded spectra, respectively.

\section{RV measurements}\label{ss:RV}  

The radial velocities were determined by cross-correlating the observed spectra from the DR1 release with theoretical template spectra calculated with the PoWR model atmosphere code \citep{Hamann+Grafener03,Sander+15}. We used template spectra from the publicly available model grids\footnote{\url{www.astro.physik.uni-potsdam.de/PoWR}} from \citet{Hamann+Grafener04,Todt+15,Hainich+19}. In total, 3610 OB and WN model spectra were used. We did not employ the WC models, as their strong emission lines make radial velocity determinations difficult.
For the SMC and LMC targets, we used the individual 1D rebinned spectra from the DR1. For each
spectrum we take the column {\tt WAVELENGTH\_AIR} for the wavelength and the column {\tt SCI\_FLUX\_NORM} for the
normalized flux. The wavelength
region over which we cross-correlate is limited to $400 - 890$~nm. The blue limit avoids the Balmer jump, while the red limit avoids the increasing level of noise.

We exhaustively compared the observed spectra to all theoretical spectra, to determine the radial velocity for each observing epoch. A grid of ten projected rotational velocities was also tried
($\varv \sin i$ = 0, 20, 50, 100, 150, 200, 250, 300, 400, 500~\kms) and all PoWR spectra were convolved with the rotational profile \citep{Gray05}.
They were also convolved with a Gaussian instrumental profile ($R=6700$ for the UVB spectrum
and $R=11\,400$ for the VIS spectrum).

For each observed target epoch, a loop is run  over all 3610 $\times$ 10 rotationally broadened PoWR spectra. Each PoWR spectrum is cross-correlated with the observed spectrum. We masked out a number of wavelength regions: the region around \heb~\l4686, the interstellar \ion{Na}{i} lines, the regions that have been corrected for telluric lines, the gap around 636~nm (which has been interpolated over in the normalized spectra),
and the Paschen P16 line (which has an incorrect wavelength in the PoWR spectra for grids computed before 2021 -- see bug warning on the PoWR website).
We further masked the pixels which have a non-zero value in the {\tt SCI\_FLUX\_QUAL} column.
When there are repeated observations of one target,  all spectra of that star are fit with the same theoretical spectrum and the same rotational broadening, but allowing for different radial velocities for each epoch. Our cross-correlation technique uses a velocity grid from $-200$ to $+600$~\kms\ with a step-size of 10~\kms. The final value for the radial velocity is
then found by a parabolic interpolation around the maximum of the cross-correlation function.

The cross-correlation function is
computed as \cite[e.g.,][]{David+14}:
\begin{equation}
C(v_s) = \frac{\sum_i (F_i - \overline{F})(T_i - \overline{T})}{\sqrt{\sum_i \left(F_i-\overline{F}\right)^2} \sqrt{\sum_i \left(T_i(v_s)-\overline{T(v_s)}\right)^2}},
\end{equation}
where $F_i$ are the observed fluxes, with the average observed flux $\overline{F}$, and $T_i(v_s)$ the theoretical fluxes, shifted by a velocity $v_s$, with the average theoretical flux $\overline{T(v_s)}$.

A parabolic fit is applied to find the maximum of the cross-correlation function, and thus the corresponding radial velocity and radial velocity uncertainty. For the uncertainty $\sigma_{\rm RV}$, we use:
\begin{equation}
\sigma_{\rm RV}^2 = - \left[N \frac{C'' C}{1-C^2}\right]^{-1},
\label{equation RV uncertainty}
\end{equation}
where $C$ is the cross-correlation function at maximum, $C''$ its second derivative,
and $N$ the number of points in the spectrum  \citep{Zucker03}.

Among all the PoWR spectra we choose the one that has the highest maximum in the cross-correlation function and use that one to determine the radial velocity.
When there are repeated observations, we chose the PoWR spectrum that has the highest sum of the cross-correlation maxima.
The resulting radial velocities are given in Table~\ref{t:RV_results}.
Table~\ref{table radial velocity excluded} lists the stars for which we could not determine the radial velocity, including the reason.

Although we list the chosen PoWR spectrum for each star in Table~\ref{t:RV_results}, 
we caution that its parameters are not necessarily the ones
that best represent the properties of the star. The radial-velocity determination procedure focuses on other
features of the spectrum than a stellar parameter determination procedure does. As one example, we remind that the
\heb~\l4686 line was excluded from the procedure. Also, most of the chosen PoWR spectra we used turn out to have SMC metallicity, even for the
LMC stars.

\begin{figure}
\centering
  \includegraphics[bb=45 28 280 270,clip, width=\columnwidth]{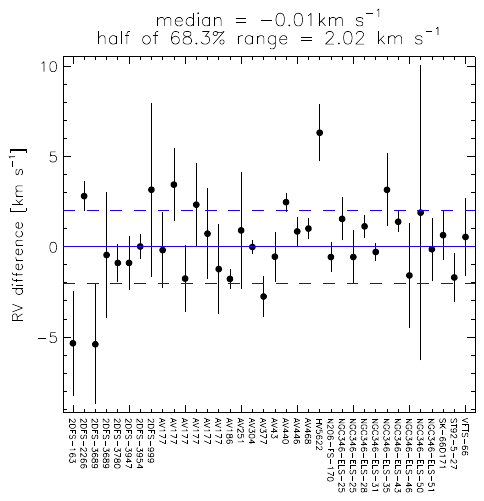}

    \caption{Radial velocity difference between observations
	of the same star (listed on the x-axis) taken 1 hour or less apart. The title lists the median and the dispersion.
 The dashed lines indicate the $\pm 2$ \kms range.}
	\label{fig radial velocity short term}
\end{figure}

The formal uncertainty derived from Eq.~\ref{equation RV uncertainty} is small for most of the stars.
As an alternative estimate of the uncertainty, we measured the short-term radial velocity behavior.
Figure~\ref{fig radial velocity short term}  shows the difference in radial velocity between observations
of the same star obtained one hour or less apart. The differences are $\sim$~2~\kms, which is a more
representative value of the radial velocity uncertainty.

As a further indication
of the reliability of our radial velocity determination, we also determined a quality index.
We introduced different quality criteria that give a weight to how
well the theoretical spectrum fits certain observed spectral lines. For the VIS spectra, this includes
judging the depth of the Paschen lines,
which play an important role in the radial velocity determination. We judged by eye
whether the observed lines are broad or narrow, and how
well the theoretical spectrum fits the observed one.
For each of these criteria we assigned points (positive or negative) and determined
the final quality indicator by adding up the points.
The details are given in Table~\ref{table RV quality}.

\begin{figure}
	\begin{center}
		\resizebox{\hsize}{!}{\includegraphics[bb=28 28 311 283,clip]{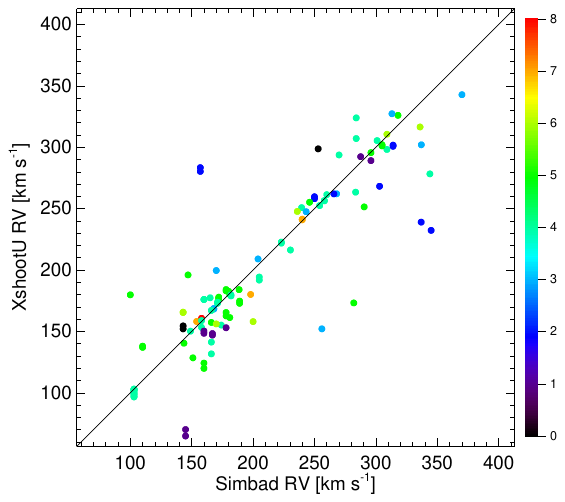}}
	\end{center}
    \caption{Comparison between our radial velocities and those from Simbad.
     The colour scale indicates the quality of our radial velocity determination.
     The formal errors on our results are, in most cases, smaller than the
     symbol.}
	\label{fig radial velocity simbad}
\end{figure}

For approximately half of our stars, 
Simbad also lists a radial velocity.
Fig.~\ref{fig radial velocity simbad} compares our values to those of Simbad. The results are colour coded
according to the quality of the RV determination. The comparison shows how the
agreement between Simbad and our results improves as we go to higher quality measurements as defined in Table~\ref{table RV quality}.

\begin{table}
        \caption{Point system for the quality indicator of the radial velocity determination.}
        \label{table RV quality}
        \begin{tabular}{llr}
                \hline\hline
Arm & Criterion & Points \\
                \hline
UVB & sharp lines & $+1$ \\
    & ... or very sharp lines & $+2$ \\
    & good fit & $+1$ \\
    & ... or very good fit  & $+2$ \\
    & emission or nebular emision & $-2$ \\
VIS & depth Paschen lines less than 0.05 & $0$ \\
    & depth Paschen lines between 0.1 and 0.2 & $+1$ \\
    & depth Paschen lines between 0.3 and 0.4 & $+2$ \\
    & good fit Paschen lines & $+1$ \\
    & good fit H$\alpha$ & $+1$ \\
    & broad lines & $-1$ \\
                \hline
        \end{tabular}
\end{table}

\section{Data release}\label{s:datarelease}  
The \xshootu\ pipeline reduced data, and the 2D and 1D data products are released to the community using [CDS, Zenodo]. A part of these products are also ingested into the ESO phase 3 archive\footnote{{\tt \url{http://www.eso.org/p3}}} as well as on the \ullyses\ website\footnote{{\tt \url{http://www.stsci.edu/ullyses}}}. The UVB/VIS data release associated to this paper is dubbed DR1. It replaces  non-public early releases (eDR) that have been distributed in the consortium for test and development purposes. We do not envision significant changes of the UVB/VIS data reduction beyond DR1, but the numbering will allow us to keep track of changes should these be required. DR2 should include the NIR but is not anticipated before the end of 2024 because the ESO pipeline is currently undergoing significant improvement of the wavelength calibration using telluric lines. In addition, the {\it HST\/} \ullyses\ target list has recently been expanded using archival data. The \xshootu\ collaboration is currently seeking to obtain the corresponding \xshooter\ observations. Should those be acquired, they will be included in a subsequent data release. 

The data format and the various data products that we provide to the community are extensively described in dedicated, publicly available documentation {\bf (Link to be added upon publication)}. We provide here a brief overview of the available content.

\subsection{Pipeline reduction products}\label{ss:PDP}
Pipeline reduction products (PRPs) are produced by running the esorex pipeline cascade (Sect.~\ref{ss:DRoverview}). We collected all intermediate PRPs and master calibrations on top of the reduced 1D and 2D spectra. For example, we obtained 20 PRP files from the reduction of each scientific frame. These PRPs are described extensively in the XSHOOTER pipeline manual that can be retrieved, e.g., from the instrument webpage\footnote{ \url{https://www.eso.org/sci/facilities/paranal/instruments/xshooter/doc.html}} and will not be repeated here. 

\subsection{Consolidated 2D data products}\label{ss:2D-DP}
We provide a subset of the most important 2D PRPs as consolidated 2D reduction products (CRP\_2D). These are formed by one set of three multi-layer fits files for each star:\begin{enumerate}
    \item [$\bullet$] {\tt XSHOOTU\_<target>\_FLUX\_MERGE2D}: flux calibrated 2D science spectra;
    \item [$\bullet$] {\tt XSHOOTU\_<target>\_FADU\_MERGE2D}: non-flux calibrated 2D science spectra;
    \item [$\bullet$] {\tt XSHOOTU\_<target>\_SKYADU\_MERGE2D}: non-flux calibrated 2D sky spectra.
\end{enumerate}
 These files provide, for each target, the reduced  (flat-fielded, wavelength calibrated, straightened, resampled and merged orders) 2D spectra corresponding to each science exposure. These 
 are produced, respectively, from the {\tt SCI\_SLIT\_FLUX\_MERGE2D\_<ARM>}, {\tt SCI\_SLIT \_MERGE2D\_<ARM>} and {\tt SKY\_SLIT \_MERGE2D\_<ARM>} PRPs. Aside from the main extension (empty, though fits descriptors allow to identify the target and DR version), each file thus contains $N_\mathrm{epoch} \times N_\mathrm{arm} \times 3$ layers, where $N_\mathrm{epoch}$ is the number of individual observing epochs and $N_\mathrm{arm}$  is the number of arm in the DR. The factor 3 come from the fact that we provide {\tt FLUX}, {\tt ERRS} and {\tt QUAL} layers for each exposure.  Each extension is tagged by the ARM, observing time and nature ({\tt FLUX}, {\tt ERRS}, {\tt QUAL}) of the data. 
 ESO headers are preserved in the headers of the individual extensions. These CRP\_2D files should be the entry points to investigate the spatial surroundings of the \xshootu\ targets (e.g., nebulosity, nearby objects in the slit), or to perform a custom-made sky subtraction or extraction of the science spectrum.

\subsection{Consolidated 1D data products}\label{ss:1D-DP}
All extracted 1D science and sky spectra (epochs, arms), and the advanced data products of each exposure are consolidated into a set of two fits files for each target:\begin{enumerate}
    \item [$\bullet$]	{\tt XSH-U\_<target>\_INDIV1D}: non-barycentric corrected 1D science and sky spectra;
    \item [$\bullet$]	{\tt XSH-U\_<target>\_INDIV1D\_REBIN}: barycentric corrected and resampled 1D science and sky spectra;
 \end{enumerate}
The {\tt INDIV1D} and {\tt INDIV1D\_REBIN} files have the same structure and contain similar information, but the wavelength calibration scale (WLC) of the {\tt INDIV1D} files is imported directly from the PRPs while the WLC of the {\tt INDIV1D\_REBIN} files has been corrected for the barycentric motion and the 1D data has been resampled onto a common WLC grid using a quadratic interpolation. This common grid starts at 300 and 545~nm for the UVB and VIS arm, respectively, and has a step of 0.02~nm, thus providing about 2 to 4 pixels per resolution elements. The WLC is provided in  air ($\lambda_\mathrm{air}$) as well as in  vacuum ($\lambda_\mathrm{vac}$) so that the data can be easily combined with the HST \ullyses\ UV data that are in the vacuum. Vacuum wavelengths were computed using the PyAstronomy\footnote{\url{https://github.com/sczesla/PyAstronomy}} \citep{pya} routine {\tt airtovac2}, which iteratively solves the equation:
\begin{equation}
    \lambda_\mathrm{vac} = n \lambda_\mathrm{air}, 
\end{equation}
where the refractive index $n$ is taken from \citet{1996ApOpt..35.1566C}:
\begin{equation}
    n=\frac{0.05792105}{238.0185-\lambda_{\mathrm{vac}}^{-2}}+\frac{0.00167917}{57.362-\lambda_{\mathrm{vac}}^{-2}} + 1,
\end{equation}
where $\lambda_{\mathrm{vac}}$ is the vacuum wavelength in $\mu$m, and assuming standard air\footnote{Standard air is at 15 $^{\mathrm{o}}$C, dry, a pressure of 101325 Pa, and a CO2 concentration of 450 ppm.}.
The flux-calibrated spectra were not resampled when moving from air to vacuum, leading to a small systematic error $\lesssim 10^{-3}$ relative to the flux level. This error is much smaller that the precision of our flux calibration.

As for the CRP\_2D, separate extensions provides access to the spectra obtained at different epochs and for the different arms. Each extension is tagged by the ARM and observing time. ESO headers of the  {\tt SCI\_SLIT\_FLUX\_MERGE1D\_<ARM>} PRPs are preserved in the headers of the individual extensions. Each extension provide a fits table containing, a.o., the response-curve corrected science spectra, the science and sky spectra in ADU (non response-corrected), as well as spectra corrected for slit-losses, telluric absorption and absolute flux-calibration. Correction functions (response-curve, wavelength-dependant slit-correction, normalisation function, etc) are not included to limit the size of the data but can be reconstructed from the different columns. 

These files containing the individual epochs are thus the entry point to study variability, as well as to access the extracted but unprocessed 1D spectra of the science target or the sky.

\subsection{Coadded 1D data products}\label{ss:1D-coadd}
Our final data products result from the co-addition of all collected 1D spectra, and the connection of UVB and VIS arm into a single spectrum. The result is provided in a single fits file per target:\begin{enumerate}
   \item [$\bullet$]	{\tt XSH-U\_<target>\_COADD1D}:  barycentric-corrected, resampled and co-added 1D science and sky spectra.
\end{enumerate}
As for the Consolidated 1D data products, the {\tt COADD1D} file provide the co-added and response-curve corrected science spectra, the co-added science and sky spectra in ADU (non response-corrected), as well as the co-added spectra corrected for slit-losses, telluric absorptions and absolute flux-calibration. In cases of multi-epoch observation, the co-addition ignores variability. The coadded 1D products should be suitable for atmosphere analysis or spectral library applications pending the target is suitable for the science goals of the analysis, e.g., in terms of variability vs. average spectrum. In particular, we do not apply any correction for RV variation of the source, such as those expected from spectroscopic binaries.

\subsection{Quality control figures and atlas} \label{ss:QCfig} 
Our data release is accompanied by a number of figures that allow the community to quickly assess the quality of the spectra, of the telluric correction and  of the normalisation. Examples of such figures are given in Figs.~\ref{f:FluxCal}, \ref{f:QCtell} and \ref{f:QCnorm} and are available on the project website\footnote{\url{http://www.massivestars.org/xshootu}} and [Zenodo]. An atlas of the flux-calibrated and normalized coadded spectra is provided in appendix (LMC: App.~\ref{app:lmc}, SMC: App.~\ref{app:smc}) available in the online version of this paper. 

\section{Summary} \label{s:ccl}
We reduced the UVB and VIS \xshooter\ data obtained by the \xshootu\ large program in a consistent and systematic way, improving upon the standard reduction in a number of ways that include the flat-fielding strategy and the use of improved {\flux} standard models. We provide advanced reduction products, including the correction for slit-losses and telluric lines, and an  absolute flux calibration. We also provide normalized spectra using smooth functions that should not have introduced high-order frequencies in the resulting spectra as the latter might modify the quantitative science content of the data.
Various data products are provided to the community, including 2D and 1D consolidated reduction products as well as  barycentric motion corrected, coadded spectra for which all individual epochs and the data from the UVB and VIS are combined to optimise the \snr.
 We also provide  quick-look figures that enable a first assessment of the quality  of various steps in the data processing as well as of the final data products.
The DR1 data should be suitable for immediate scientific exploitation.

\section*{Acknowledgments}
we thank the referee for their constructive comments. 
This project has received funding from the KU Leuven Research Council (grant C16/17/007: MAESTRO).
M.A. acknowledges support from ”la Caixa” Foundation (ID 100010434) under the fellowship code LCF/BQ/PI23/11970035.
L.M. thanks FAPESP (grant 2022/03703-1) and CNPQ (grant
307115/2021-6) for partial funding of this research.
AACS and VR are supported by the Deutsche Forschungsgemeinschaft (DFG - German Research Foundation) in the form of an Emmy Noether Research Group -- Project-ID 445674056 (SA4064/1-1, PI Sander). AACS and VR are further supported by funding from the Federal Ministry of Education and Research (BMBF) and the Baden-Württemberg Ministry of Science as part of the Excellence Strategy of the German Federal and State Governments.
JSV gratefully acknowledges support from STFC via grant ST/V000233/1. CJKL gratefully acknowledges support from the International Max Planck Research School for Astronomy and Cosmic Physics at the University of Heidelberg in the form of an IMPRS PhD fellowship. GM acknowledges funding support from the European Research Council (ERC) under the European Union’s Horizon 2020 research and innovation programme (Grant agreement No. 772086). F.N. acknowledges funding by grants PID2019-105552RB-C41 and MDM-
2017-0737-19-3 Unidad de Excelencia "María de Maeztu".
CK acknowledges financial support from the Spanish Ministerio de Econom\'{\i}a y Competitividad under grants AYA2016-79724-C4-4-P and PID2019-107408GB-C44, from Junta de Andaluc\'{\i}a Excellence Project P18-FR-2664, and from the grant CEX2021-001131-S funded by MCIN/AEI/ 10.13039/501100011033. PAC and JB are supported by the Science and Technology Facilities Council research grant ST/V000853/1 (PI. V. Dhillon). RK acknowledges financial support via the Heisenberg Research Grant funded by the German Research Foundation (DFG) under grant no.~KU 2849/9.

\bibliographystyle{aa} 
\bibliography{reference.bib} 

\begin{appendix}
	\onecolumn
\section{Radial velocity results}

        \begin{longtable}{rlrrrlrrl}
                \caption{Radial velocity results.}\\
        \label{t:RV_results}\\
                \hline
                \hline
no. & \multicolumn{1}{c}{Object} & \multicolumn{1}{c}{MJD} & RV & Err& \multicolumn{1}{c}{Template} & $\varv \sin i$ & qual & Comment\\
&      &  & (\kms) & (\kms) & \multicolumn{1}{c}{PoWR model} & (\kms) & qual & comment\\
                \hline
                \endfirsthead
                \caption{continued}\\
                \hline
no. & \multicolumn{1}{c}{Object} & \multicolumn{1}{c}{MJD} & RV & Err& \multicolumn{1}{c}{Template} & $\varv \sin i$ & qual & Comment\\
&      &  & RV & Err & \multicolumn{1}{c}{PoWR model} & $\varv \sin i$ & qual & comment\\
                \hline
                \endhead
                \hline
                \endfoot
                \hline
                \endlastfoot
  1 &                         AV~14 & 59156.695 &  144.1 &   0.8 &                          smc-ob-i\_43-40 &  150 &    5 &          XB (Simbad) \\
  2 &                         AV~15 & 59155.047 &  128.5 &   0.5 &                         smc-ob-ii\_38-36 &  100 &    5 &                      \\
  3 &                         AV~18 & 59155.058 &  152.2 &   0.4 &                        smc-ob-iii\_20-26 &   50 &    6 &                      \\
  4 &                         AV~22 & 59155.012 &  140.4 &   0.3 &                          smc-ob-i\_16-22 &   50 &    5 &                      \\
  5 &                         AV~43 & 59157.252 &  168.7 &   1.0 &                          smc-ob-i\_24-32 &  250 &    3 &                      \\
    &                                & 59157.261 &  168.2 &   1.0 &                          smc-ob-i\_24-32 &  250 &    3 &                      \\
... & ... & ... & ... & ... & ... & ... & ... & ... \\
        \end{longtable}
\tablefoot{
Column (1) Running number or the star;
(2) Star name -- a given star can have multiple entries if there are multiple observations more than 15 minutes apart;
(3) Modified Julian date of the observation (start of observation) minus 2,450,000.0;
(4) Radial velocity;
(5) Uncertainty on radial velocity – an underestimate in most cases;
(6) Code of the template PoWR model used;
(7) Projected rotational velocity used in the cross-correlation;
(8) Quality of the fit (range 0 -- 10), where low values indicate a bad quality and high values a good quality;
(9) comments about possible binarity, mostly from Simbad. EB=eclipsing binary, XB=X-ray binary.
The full table is available at CDS.
}

\begin{table}[h]
\centering
        \caption{Stars for which we do not provide a radial velocity.}
        \label{table radial velocity excluded}
        \begin{tabular}{ll}
                \hline\hline
                Star & Comment \\
                \hline
2DFS-163  &   binary  \\
AV~83 & bad fit + emission UVB arm \\
AV~189 & SB2?  \\
NGC346-ELS-051 & bad fit + nebular emission UVB arm \\
2DFS-2553 & SB2  \\
2DFS-3689 & eclipsing binary (Simbad); SB2  \\
SK-173 & emission \\
PGMW-3070 & bad fit + emission VIS arm \\
LH-58-496 &  bad fit  \\
SK-67D118 & SB2?  \\
SK-71D35 & emission \\
                \hline
        \end{tabular}
\end{table}

\section{Individual notes}

\subsection*{AV~324} 
Observations in poor seeing conditions (up to 2\arcsec). We adopted a broader extraction window for the central objects and  smaller windows for sky regions.

\subsection*{BAT99 105}   
The faint companion on the lower part pf the slit has no impact on the sky estinate, probably because of the median approach used in th pipeline to  construct the sky spectrum.

\subsection*{HD269927C ($\equiv$ BAT99 120)} 
We note a very faint companion in th elower part of the slit, contributing  less that 1\%\ of the flux of the central object. We also note variable nebulosity across the slit. The standard extraction windows, where the companion is included in teh sky regions, result in a better nebular subtraction that when only using the upper sky region ({\it Default Sky 2} in Fig.~\ref{f:sky}). We opted for the former approach albeit  the companion probably contribute to $\sim0.5$\%\ of the extracted flux. The shape of the Balmer lines (strong wind lines) in the extracted spectrum of HD269927C do not seem to different in either approach. 

\subsection*{SK-67D106} 
Observations in poor seeing conditions (up to 2\arcsec). We adopted a broader extraction window for the central objects and  smaller windows for sky regions.

\subsection*{SK-191} 
Observations in poor seeing conditions (up to 2\arcsec). We adopted a broader extraction window for the central objects and  smaller windows for sky regions.

\subsection*{VFTS-482}
The slit contains at least three other objects, at approx. position at -2\farcs4, $+$2\farcs2 and $+$5\farcs0. Only a $\approx$1\farcs2 window towards the bottom of the slit is available for the estimate of the sky background. The brightest of the three companions is too close to the edge of the slit for a meaningful extraction while the other two are  too faint.

\subsection*{VFTS-506}
Both edges of the slit seem contaminated by nearby sources, but this is not affecting the 2D sky subtracted spectra thanks to the limited contamination and the median approach used. The sky level seem larger than for other stars in the 30~Dor region, possibly due to nearby faint companions that are not clearly resolved. The level of contamination, if present, is likely faint or even negligible so that we used the start extraction window for VFTS~506.

\subsection*{VFTS-542}
 A nearby visual companion is identified at +4.5\arcsec\ in the upper part of the slit. Another much fainter companion is also present at +2.2\arcsec so that we do not use the upper part of the slit to estimate the background/sky. Finally, there is a possible very faint companion at about $-$3\arcsec, i.e.\ within the adopted sky region. Its trace is  only visible in the reddest part of the VIS arm on raw image and its contribution is probably completely negligible, so that we have ignored it. 
 
 The companion at +4.5\arcsec\ presents clear signature of Balmer lines, and of \hea~\l\l4443, 4471 and \heb~\l4686 absorption. 

\section{LMC atlas} \label{app:lmc}

\section{SMC atlas} \label{app:smc}

\section{Target names} \label{app:targets}
\begin{longtable}{l l l l}
\caption{Overview of target names used in XShootU, MAST, and Simbad.}
\label{t:names}\\
\hline\hline
XShootU$^a$ & OB$^b$ & MAST$^c$ & Simbad \\
\hline \\[-9pt]
\endfirsthead
\caption{continued.} \\
\hline \hline
XShootU$^a$ & OB$^b$ & MAST$^c$ & Simbad \\
\hline \\[-9pt]
\endhead
\hline
\multicolumn{4}{l}{
\footnotesize{{\bf Notes. }$^{(a)}$ Target name from Table B.1 of \citetalias{xsh1}. $^{(b)}$ OB name, as stored in the OBJECT keywords of the RAW data files.}}\\
\multicolumn{4}{l}{
\footnotesize{$^{(c)}$ MAST identifier, used for filenames and OBJECT keywords of the HLSPs.}}
\endfoot
\multicolumn{4}{c}{{\bf SMC Targets}} \\
2dFS 163	& 2dFS-163	& 2DFS-163	& \object{2dFS 163}\\
AzV 6	& AzV-6	& AV-6	& \object{AzV 6}\\
AzV 14	& AzV-14	& AV-14	& \object{AzV 14}\\
AzV 15	& AzV-15	& AV-15	& \object{AzV 15}\\
AzV 16	& AzV-16	& AV-16	& \object{AzV 16}\\
AzV 18	& AzV-18	& AV-18	& \object{AzV 18}\\
AzV 22	& AzV-22	& AV-22	& \object{AzV 22}\\
AzV 39a	& AzV-39a	& AV-39A	& \object{AzV 39a}\\
AzV 43	& AzV-43	& AV-43	& \object{AzV 43}\\
AzV 47	& AzV-47	& AV-47	& \object{AzV 47}\\
AzV 61	& AzV-61	& AV-61	& \object{AzV 61}\\
AzV 69	& AzV-69	& AV-69	& \object{AzV 69}\\
AzV 70	& AzV-70	& AV-70	& \object{AzV 70}\\
AzV 75	& AzV-75	& AV-75	& \object{AzV 75}\\
AzV 81	& SK-41	& AV-81	& \object{SK 41}\\
AzV 80	& AzV-80	& AV-80	& \object{AzV 80}\\
AzV 83	& AzV-83	& AV-83	& \object{AzV 83}\\
AzV 85	& AzV-85	& AV-85	& \object{AzV 85}\\
AzV 95	& AzV-95	& AV-95	& \object{AzV 95}\\
AzV 96	& AzV-96	& AV-96	& \object{AzV 96}\\
AzV 104	& AzV-104	& AV-104	& \object{AzV 104}\\
AzV 148	& AzV-148	& AV-148	& \object{AzV 148}\\
2dFS 999	& 2dFS-999	& 2DFS-999	& \object{SMC AB 9}\\
NGC330 ELS 4	& 2dFS-5090	& NGC330-ELS-004	& \object{CL* NGC 330 ELS 4}\\
NGC330 ELS 2	& Gaia-DR2-	& NGC330-ELS-002	& \object{CL* NGC 330 ELS 2}\\
                & \hspace*{3mm}4688993475604539520 \\
AzV 175	& AzV-175	& AV-175	& \object{AzV 175}\\
AzV 177	& AzV-177	& AV-177	& \object{AzV 177}\\
AzV 186	& AzV-186	& AV-186	& \object{Cl* NGC 330 ELS 013}\\
AzV 189	& AzV-189	& AV-189	& \object{AzV 189}\\
AzV 200	& AzV-200	& AV-200	& \object{AzV 200}\\
NGC346 ELS 43	& Gaia-DR2-	& NGC346-ELS-043	& \object{CL* NGC 346 ELS 43}\\
 & \hspace*{3mm}4689019245413731840\\
NGC346 ELS 26	& 2dFS-1299	& NGC346-ELS-026	& \object{CL* NGC 346 ELS 26}\\
NGC346 ELS 28	& 2MASS-J00583173-7210580	& NGC346-ELS-028	& \object{Cl* NGC 346 ELS 028}\\
AzV 207	& AzV-207	& AV-207	& \object{AzV 207}\\
AzV 210	& AzV-210	& AV-210	& \object{AzV 210}\\
NGC346 ELS 50	& Gaia-DR2-	& NGC346-ELS-050	& \object{Cl* NGC 346 ELS 050}\\
& \hspace*{3mm}4689016530994364672\\
AzV 215	& AzV-215	& AV-215	& \object{AzV 215}\\
NGC346 ELS 7	& 2MASS-J00585738-7210336	& NGC346-ELS-007	& \object{CL* NGC 346 ELS 7}\\
NGC346 MPG 355	& 2MASS-J00590075-7210282	& NGC346-MPG-355	& \object{Cl* NGC 346 MPG 355}\\
NGC346 MPG 368	& NGC-346-MPG-368	& NGC346-MPG-368	& \object{Cl* NGC 346 MPG 368}\\
NGC346 MPG 435	& 2MASS-J00590446-7210248	& NGC346-MPG-435	& \object{Cl* NGC 346 W 1}\\
NGC346 ELS 51	& 2MASS-J00590866-7210142	& NGC346-ELS-051	& \object{Cl* NGC 346 ELS 051}\\
AzV 224	& AzV-224	& AV-224	& \object{AzV 224}\\
HD 5980	& HD-5980	& HD-5980	& \object{HD 5980}\\
NGC346 ELS 13	& OGLE-SMC108.3-14355	& NGC346-ELS-013	& \object{Cl* NGC 346 ELS 013}\\
NGC346 ELS 46	& 2MASS-J00593188-7213352	& NGC346-ELS-046	& \object{Cl* NGC 346 ELS 046}\\
AzV 232	& AzV-232	& AV-232	& \object{AzV 232}\\
AzV 234	& AzV-234	& AV-234	& \object{AzV 234}\\
AzV 235	& AzV-235	& AV-235	& \object{SK 82}\\
NGC346 ELS 35	& 2dFS-1418	& NGC346-ELS-035	& \object{Cl* NGC 346 ELS 035}\\
NGC346 ELS 25	& Gaia-DR2-	& NGC346-ELS-025	& \object{Cl* NGC 346 ELS 025}\\
& \hspace*{3mm}4690504994818102016\\
NGC346 ELS 31	& 2MASS-J00595407-7204316	& NGC346-ELS-031	& \object{Cl* NGC 346 ELS 031}\\
AzV 238	& AzV-238	& AV-238	& \object{AzV 238}\\
AzV 243	& AzV-243	& AV-243	& \object{AzV 243}\\
AzV 242	& AzV-242	& AV-242	& \object{AzV 242}\\
AzV 251	& AzV-251	& AV-251	& \object{AzV 251}\\
AzV 255	& AzV255	& AV-255	& \object{AzV 255}\\
AzV 264 & AzV-264   & AV-264 & \object{AzV 264}\\
AzV 266	& AzV-266	& AV-266	& \object{AzV 266}\\
AzV 267	& AzV-267	& AV-267	& \object{AzV 267}\\
AzV 296	& AzV-296	& AV-296	& \object{AzV 296}\\
AzV 304	& AzV-304	& AV-304	& \object{AzV 304}\\
AzV 307	& AzV-307	& AV-307	& \object{AzV 307}\\
AzV 314	& AzV-314	& AV-314	& \object{AzV 314}\\
AzV 321	& AzV-321	& AV-321	& \object{AzV 321}\\
AzV 324	& AzV-324	& AV-324	& \object{AzV 324}\\
AzV 326	& AzV-326	& AV-326	& \object{AzV 326}\\
AzV 327	& AzV-327	& AV-327	& \object{AzV 327}\\
AzV 332	& SK-108	& AV-332	& \object{SK 108}\\
AzV 343	& AzV-343	& AV-343	& \object{AzV 343}\\
AzV 362	& AzV-362	& AV-362	& \object{AzV 362}\\
AzV 372	& AzV-372	& AV-372	& \object{AzV 372}\\
AzV 374	& AzV-374	& AV-374	& \object{AzV 374}\\
AzV 377	& AzV-377	& AV-377	& \object{AzV 377}\\
AzV 388	& AzV-388	& AV-388	& \object{AzV 388}\\
AzV 393	& SK-124	& AV-393	& \object{SK 124}\\
AzV 410	& AzV-410	& AV-410	& \object{AzV 410}\\
2dFS 2266	& 2dFS-2266	& 2DFS-2266	& \object{2dFS 2266}\\
AzV 423	& AzV-423	& AV-423	& \object{AzV 423}\\
AzV 435	& AzV-435	& AV-435	& \object{AzV 435}\\
AzV 440	& AzV-440	& AV-440	& \object{AzV 440}\\
AzV 445	& AzV-445	& AV-445	& \object{AzV 445}\\
2dFS 2553	& 2dFS-2553	& 2DFS-2553	& \object{2dFS 2553}\\
AzV 446	& AzV-446	& AV-446	& \object{AzV 446}\\
AzV 456	& AzV-456	& AV-456	& \object{AzV 456}\\
AzV 468	& AzV-468	& AV-468	& \object{AzV 468}\\
AzV 469	& AzV-469	& AV-469	& \object{AzV 469}\\
AzV 476	& AzV-476	& AV-476	& \object{AzV 476}\\
AzV 479	& AzV-479	& AV-479	& \object{AzV 479}\\
AzV 488	& AzV-488	& AV-488	& \object{SK 159}\\
AzV 490	& AzV-490	& AV-490	& \object{AzV 490}\\
AzV 506	& AzV-506	& AV-506	& \object{AzV 506}\\
$[$M2002$]$ SMC 81469	& Gaia-DR2-	& M2002-SMC-81469	& \object{[M2002] SMC 81469}\\
& \hspace*{3mm}4686450580031889536\\
2dFS 3689	& 2dFS-3689	& 2DFS-3689	& \object{2dFS 3689}\\
2dFS 3694	& 2dFS-3694	& 2DFS-3694	& \object{2dFS 3694}\\
SK 173	& SK-173	& SK-173	& \object{2dFS 3747}\\
2dFS 3780	& 2dFS-3780	& 2DFS-3780	& \object{2dFS 3780}\\
SK 179	& SK-179	& SK-179	& \object{SK 179}\\
SK 183	& SK-183	& SK-183	& \object{SK 183}\\
2dFS 3947	& 2dFS-3947	& 2DFS-3947	& \object{2dFS 3947}\\
2dFS 3954	& 2dFS-3954	& 2DFS-3954	& \object{2dFS 3954}\\
SK 187	& SK-187	& SK-187	& \object{SK 187}\\
SK 191	& SK-191	& SK-191	& \object{SK 191}\\
\hline \\[-9pt]
\multicolumn{4}{c}{{\bf LMC Targets}} \\
SK -67\degr\ 2	& SK-\,-67-02	& SK-67D2	& \object{SK -67 02
}\\
SK -67\degr\ 5	& SK-\,-67-05	& SK-67D5	& \object{SK -67 05
}\\
BI 13	& BI-13	& BI-13	& \object{BI 13
}\\
SK -68\degr\ 8	& SK-\,-68-8	& SK-68D8	& \object{SK -68 8
}\\
SK -70\degr\ 13	& SK-\,-70-13	& SK-70D13	& \object{SK -70 13
}\\
SK -67\degr\ 14	& SK-\,-67-14	& SK-67D14	& \object{SK -67 14
}\\
SK -70\degr\ 16	& SK-\,-70-16	& SK-70D16	& \object{SK -70 16
}\\
SK -67\degr\ 20	& HD-32109	& SK-67D20	& \object{HD 32109
}\\
SK -66\degr\ 17	& SK-\,-66-17	& SK-66D17	& \object{[ELS2006] N11 011
}\\
SK -66\degr\ 18	& SK-\,-66-18	& SK-66D18	& \object{SK -66 18
}\\
SK -69\degr\ 43	& SK-\,-69-43	& SK-69D43	& \object{SK -69 43
}\\
N11 ELS 33	& N11-033	& N11-ELS-033	& \object{[ELS2006] N11 033
}\\
N11 ELS 49	& N11-049	& N11-ELS-049	& \object{[ELS2006] N11 049
}\\
N11 ELS 51	& N11-051	& N11-ELS-051	& \object{[ELS2006] N11 051
}\\
N11 ELS 18	& PGMW-3053	& N11-ELS-018	& \object{[ELS2006] N11 018
}\\
N11 ELS 60	& N11-060	& N11-ELS-060	& \object{[ELS2006] N11 060
}\\
PGMW 3070	& PGMW-3070	& PGMW-3070	& \object{PGMW 3070
}\\
N11 ELS 46	& N11-046	& N11-ELS-046	& \object{[ELS2006] N11 046
}\\
N11 ELS 38	& N11-038	& N11-ELS-038	& \object{[ELS2006] N11 038
}\\
PGWM 1363	& PGMW-1363	& PGMW-1363	& \object{PGMW 1363
}\\
PGMW 3120	& PGMW-3120	& PGMW-3120	& \object{PGMW 3120
}\\
LMCe055-1	& OGLE-LMC-ECL-3548	& LMCE055-1	& \object{[MNM2015] LMCe055-1
}\\
N11 ELS 20	& N11-020	& N11-ELS-020	& \object{[ELS2006] N11 020
}\\
SK -65\degr\ 2	& SK-\,-65-2	& SK-65D2	& \object{SK -65 2
}\\
N11 ELS 32	& PGMW-3168	& N11-ELS-032	& \object{[ELS2006] N11 032
}\\
N11 ELS 48	& PGMW-3204	& N11-ELS-048	& \object{[ELS2006] N11 048
}\\
N11 ELS 13	& BI-42	& N11-ELS-013	& \object{[ELS2006] N11 013
}\\
SK -66\degr\ 35	& SK-\,-66-35	& SK-66D35	& \object{SK -66 35
}\\
SK -69\degr\ 50	& SK-\,-69-50	& SK-69D50	& \object{SK -69 50
}\\
SK -68\degr\ 15	& BAT99-11	& SK-68D15	& \object{BAT99 11
}\\
SK -67\degr\ 22	& SK-\,-67-22	& SK-67D22	& \object{SK -67 22
}\\
SK -68\degr\ 16	& SK-\,-68-16	& SK-68D16	& \object{SK -68 16
}\\
SK -69\degr\ 52	& SK-\,-69-52	& SK-69D52	& \object{SK -69 52
}\\
SK -70\degr\ 32	& SK-\,-70-32	& SK-70D32	& \object{SK -70 32
}\\
SK -68\degr\ 23a	& SK-\,-68-23A	& SK-68D23A	& \object{SK -68 23A
}\\
SK -65\degr\ 22	& SK-\,-65-22	& SK-65D22	& \object{SK -65 22
}\\
SK -68\degr\ 26	& SK-\,-68-26	& SK-68D26	& \object{SK -68 26
}\\
SK -66\degr\ 50	& SK-\,-66-50	& SK-66D50	& \object{SK -66 50
}\\
SK -70\degr\ 50	& SK-\,-70-50	& SK-70D50	& \object{SK -70 50
}\\
SK -70\degr\ 60	& SK-\,-70-60	& SK-70D60	& \object{SK -70 60
}\\
SK -70\degr\ 69	& SK-\,-70-69	& SK-70D69	& \object{SK -70 69
}\\
SK -68\degr\ 41	& SK-\,-68-41	& SK-68D41	& \object{SK -68 41
}\\
SK -70\degr\ 79	& SK-\,-70-79	& SK-70D79	& \object{SK -70 79
}\\
SK -68\degr\ 52	& SK-\,-68-52	& SK-68D52	& \object{SK -68 52
}\\
SK -71\degr\ 8	& SK-\,-71-8	& SK-71D8	& \object{SK -71 8
}\\
HV 5622	& MACHO-79.4779.34	& HV-5622	& \object{MACHO 79.4779.34
}\\
SK -67\degr\ 51	& SK-\,-67-51	& SK-67D51	& \object{SK -67 51
}\\
SK -67\degr\ 69	& SK-\,-67-69	& SK-67D69	& \object{SK -67 69
}\\
SK -69\degr\ 83	& SK-\,-69-83	& SK-69D83	& \object{SK -69 83
}\\
BI 128	& BI-128	& BI-128	& \object{BI 128
}\\
SK -69\degr\ 104	& SK-\,-69-104	& SK-69D104	& \object{SK -69 104
}\\
SK -67\degr\ 78	& SK-\,-67-78	& SK-67D78	& \object{SK -67 78
}\\
SK -65\degr\ 47	& SK-\,-65-47	& SK-65D47	& \object{SK -65 47
}\\
SK -65\degr\ 55	& BAT99-30	& SK-65D55	& \object{BAT99 30
}\\
SK -71\degr\ 19	& SK-\,-71-19	& SK-71D19	& \object{SK -71 19
}\\
SK -71\degr\ 21	& BAT99-32	& SK-71D21	& \object{BAT99 32
}\\
SK -68\degr\ 73	& SK-\,-68-73	& SK-68D73	& \object{SK -68 73
}\\
SK -67\degr\ 101	& SK-\,-67-101	& SK-67D101	& \object{SK -67 101
}\\
SK -67\degr\ 105	& SK-\,-67-105	& SK-67D105	& \object{SK -67 105
}\\
SK -67\degr\ 106	& SK-\,-67-106	& SK-67D106	& \object{SK -67 106
}\\
SK -67\degr\ 107	& SK-\,-67-107	& SK-67D107	& \object{SK -67 107
}\\
SK -67\degr\ 108	& SK-\,-67-108	& SK-67D108	& \object{SK -67 108
}\\
LH 58-496	& LH-58-496	& LH-58-496	& \object{[L72] LH 58-496
}\\
SK -67\degr\ 111	& SK-\,-67-111	& SK-67D111	& \object{SK -67 111
}\\
BI 173	& BI-173	& BI-173	& \object{BI 173
}\\
SK -67\degr\ 118	& SK-\,-67-118	& SK-67D118	& \object{SK -67 118
}\\
SK -69\degr\ 140	& SK-\,-69-140	& SK-69D140	& \object{SK -69 140
}\\
SK -66\degr\ 100	& SK-\,-66-100	& SK-66D100	& \object{SK -66 100
}\\
SK -71\degr\ 35	& SK-\,-71-35	& SK-71D35	& \object{SK -71 35
}\\
BI 184	& BI-184	& BI-184	& \object{BI 184
}\\
SK -71\degr\ 41	& SK-\,-71-41	& SK-71D41	& \object{SK -71 41
}\\\degr\
NGC2004 ELS 3	& RMC-109	& NGC2004-ELS-003	& \object{Cl* NGC 2004 ELS 3
}\\
BI 189	& BI-189	& BI-189	& \object{BI 189
}\\
N206-FS 170	& Gaia-DR2-	& N206-FS-170	& \object{[RHH2018] 170
}\\
& \hspace*{3mm}4657849288998781440\\
SK -71\degr\ 45	& SK-\,-71-45	& SK-71D45	& \object{SK -71 45
}\\
SK -67\degr\ 166	& SK-\,-67-166	& SK-67D166	& \object{SK -67 166
}\\
SK -71\degr\ 46	& SK-\,-71-46	& SK-71D46	& \object{SK -71 46
}\\
SK -67\degr\ 167	& SK-\,-67-167	& SK-67D167	& \object{SK -67 167
}\\
SK -67\degr\ 168	& SK-\,-67-168	& SK-67D168	& \object{SK -67 168
}\\
SK -69\degr\ 178	& SK-\,-69-178	& SK-69D178	& \object{SK -69 178
}\\
LMC X-4	& X-LMC-X-4	& LMCX-4	& \object{X LMC X-4
}\\
SK -67\degr\ 191	& SK-\,-67-191	& SK-67D191	& \object{SK -67 191
}\\
SK -67\degr\ 195	& SK-\,-67-195	& SK-67D195	& \object{SK -67 195
}\\
SK -67\degr\ 197	& SK-\,-67-197	& SK-67D197	& \object{SK -67 197
}\\
SK -66\degr\ 152	& SK-\,-66-152	& SK-66D152	& \object{SK -66 152
}\\
SK -69\degr\ 191	& BAT99-61	& SK-69D191	& \object{BAT99 61
}\\
W61 28-5	& W61-28-5	& W61-28-5	& \object{W61 28-5
}\\
W61 28-23	& W61-28-23	& W61-28-23	& \object{W61 28-23
}\\
SK -67\degr\ 207	& SK-\,-67-207	& SK-67D207	& \object{SK -67 207
}\\
SK -67\degr\ 211	& SK-\,-67-211	& SK-67D211	& \object{SK -67 211
}\\
MCPS 083.91120-69.69685	& MCPS-083.91120-69.69685	& GAIA-DR3-	& \object{MCPS 083.91120-69.69685
}\\
& & \hspace*{3mm}4657277616048732544\\
SK -67\degr\ 216	& SK-\,-67-216	& SK-67D216	& \object{SK -67 216
}\\
SK -69\degr\ 212	& SK-\,-69-212	& SK-69D212	& \object{SK -69 212
}\\
BI 237	& BI-237	& BI-237	& \object{BI 237
}\\
SK -68\degr\ 133	& SK-\,-68-133	& SK-68D133	& \object{SK -68 133
}\\
SK -66\degr\ 171	& SK-\,-66-171	& SK-66D171	& \object{SK -66 171
}\\
LMCe078-1	& MCPS-084.37367-69.24772	& LMCE078-1	& \object{[MNM2015] LMCe078-1
}\\
VFTS 66	& VFTS-66	& VFTS-66	& \object{VFTS 66
}\\
VFTS 72	& VFTS-72	& VFTS-72	& \object{BI 253
}\\
SK -68\degr\ 135	& SK-\,-68-135	& SK-68D135	& \object{SK -68 135
}\\
VFTS 169	& VFTS-169	& VFTS-169	& \object{VFTS 169
}\\
VFTS 180	& BAT99-93	& VFTS-180	& \object{VFTS180
}\\
VFTS 190	& VFTS-190	& VFTS-190	& \object{VFTS 190
}\\
VFTS 244	& VFTS-244	& VFTS-244	& \object{VFTS 244
}\\
VFTS 267	& VFTS-267	& VFTS-267	& \object{VFTS 267
}\\
SK -68\degr\ 137	& SK-\,-68-137	& SK-68D137	& \object{SK -68 137
}\\
VFTS 355	& VFTS-355	& VFTS-355	& \object{VFTS 355
}\\
VFTS 404	& VFTS-404	& VFTS-404	& \object{VFTS 404
}\\
VFTS 406	& VFTS-406	& VFTS-406	& \object{VFTS 406
}\\
VFTS 482	& VFTS-482	& VFTS-482	& \object{Cl* NGC 2070 MEL 39
}\\
VFTS 506	& VFTS-506	& VFTS-506	& \object{VFTS 506
}\\
BAT99 105	& BAT99-105	& BAT99-105	& \object{Cl* NGC 2070 MEL 42
}\\
VFTS 542	& VFTS-542	& VFTS-542	& \object{Cl* NGC 2070 MEL 30
}\\
VFTS 545	& VFTS-545	& VFTS-545	& \object{Cl* NGC 2070 MEL 35
}\\
VFTS 586	& VFTS-586	& VFTS-586	& \object{VFTS 586
}\\
SK -68\degr\ 140	& SK-\,-68-140	& SK-68D140	& \object{SK -68 140
}\\
HD 269927C	& BAT99-120	& HD-269927C	& \object{[ST92] 5-68
}\\
$[$ST92$]$ 5-52	& W61-3-14	& ST92-5-52	& \object{[ST92] 5-52
}\\
$[$ST92$]$ 5-27	& W61-3-24	& ST92-5-27	& \object{[ST92] 5-27
}\\
$[$ST92$]$ 4-18	& W61-4-4	& ST92-4-18	& \object{W61 4-4
}\\
BI 265	& BI-265	& BI-265	& \object{BI 265
}\\
Farina  88	& MCPS-085.03420-69.65476	& FARINA-88	& \object{[FBM2009] 88
}\\
SK -71\degr\ 50	& SK-\,-71-50	& SK-71D50	& \object{SK -71 50
}\\
SK -69\degr\ 279	& SK-\,-69-279	& SK-69D279	& \object{SK -69 279
}\\
SK -68\degr\ 155	& SK-\,-68-155	& SK-68D155	& \object{SK -68 155
}\\
LH 114-7	& LH-114-7	& LH-114-7	& \object{[L72] LH 114-7
}\\
BI 272	& BI-272	& BI-272	& \object{BI 272
}\\
SK -67\degr\ 261	& SK-\,-67-261	& SK-67D261	& \object{SK -67 261
}\\
SK -70\degr\ 115	& SK-\,-70-115	& SK-70D115	& \object{SK -70 115}\\
\end{longtable}

\end{appendix}

\end{document}